\documentclass[11pt]{report}
\usepackage{bm}
\usepackage{dsfont}
\usepackage{xcolor}
\usepackage{varwidth}
\usepackage{array}
\usepackage[top=0.5cm, bottom=2.5cm]{geometry}
\usepackage[bookmarks=true,bookmarksdepth=4]{hyperref}
\usepackage{transparent}
\usepackage{pdfpages}
\usepackage{amsmath}
\usepackage{amssymb}
\usepackage{amsfonts}
\usepackage{graphicx}
\usepackage{amsthm}
\usepackage{tikz-cd}
\usepackage{lipsum}
\usetikzlibrary{cd}
\usepackage{tikz}
\usepackage{stackengine}
\usepackage{appendix}
\usepackage{mathrsfs}
\usepackage{eso-pic}
\usepackage{titlesec, blindtext, color}
\usepackage[T1]{fontenc}
\usepackage{etoolbox}
\usepackage{multicol}
\usepackage{wrapfig}
\usetikzlibrary{positioning,backgrounds,decorations.markings}
\usepackage{marginnote}
\usepackage{afterpage}
\usepackage[framemethod=tikz]{mdframed}
\hypersetup{
	colorlinks=true,
	linkcolor=bluegreen,
	urlcolor=black,
	citecolor=bluegreen,
	filecolor=bluegreen,
	bookmarks=true,}
\tikzset{>=stealth}

\def\fauxschelper#1 #2\relax{%
	\fauxschelphelp#1\relax\relax%
	\if\relax#2\relax\else\ \fauxschelper#2\relax\fi%
}
\def\Hscale{.85}\def\Vscale{.75}\def\Cscale{1.10}
\def\fauxschelphelp#1#2\relax{%
	\ifnum`#1>``\ifnum`#1<`\{\scalebox{\Hscale}[\Vscale]{\uppercase{#1}}\else%
	\scalebox{\Cscale}[1]{#1}\fi\else\scalebox{\Cscale}[1]{#1}\fi%
	\ifx\relax#2\relax\else\fauxschelphelp#2\relax\fi}
\def\theLetterSpace{0.5pt}
\def\theWordSpace{1\LocalLetterSpace}
\newlength\LocalLetterSpace
\newcommand\spaceout[2][\theLetterSpace]{%
	\setlength\LocalLetterSpace{#1}\expandafter\spaceouthelpA#2 \relax\relax}
\def\spaceouthelpA#1 #2\relax{%
	\spaceouthelpB#1\relax\relax%
	\ifx\relax#2\else\ \kern\theWordSpace\spaceouthelpA#2\relax\fi
}
\def\spaceouthelpB#1#2\relax{%
	#1%
	\ifx\relax#2\else
	\kern\LocalLetterSpace\spaceouthelpB#2\relax%
	\fi
}
\usepackage[some]{background}

\definecolor{lightgrey}{rgb}{.9,.9,.9}
\definecolor{darkgrey}{rgb}{.05,.05,.05}
\definecolor{DarkGray}{RGB}{0,0,0}
\makeatletter
\newcommand\HUGE{\@setfontsize\Huge{28}{0}}
\makeatother  
\RequirePackage{fix-cm}
{
	\definecolor{bluegreen}{rgb}{-.2,.4,0.6}
	\definecolor{lightblue}{HTML}{5dbcd2}
	\makeatletter

	\def\@seccntformat#1{\csname the#1\endcsname\hspace*{0.5em}$|$\hspace*{0.5em}}
	\makeatother
	\definecolor{gray55}{gray}{0.55}
	\newcommand{\hsp}{\hspace{2.5pt}}

	\setcounter{secnumdepth}{4}
	
	\titleformat{\chapter}[hang] {\vspace{-3em}\LARGE\scshape\bfseries}{\hspace{-.5cm} \thechapter\hsp\hsp\hsp{|}\hsp\hsp\hsp}{0pt}{\LARGE\bfseries}
	\titleformat*{\section}{\large\bfseries\scshape}
	\titleformat*{\subsection}{\large\bfseries\scshape}
	\titleformat*{\subsubsection}{\normalsize\bfseries\scshape}
	\titlespacing*{\chapter}{0pt}{25pt}{100pt}
	\titlespacing*{\section}{0pt}{15pt}{15pt}
	\titlespacing*{\subsection}{0pt}{12.5pt}{12.5pt}
	\titlespacing*{\subsubsection}{0pt}{10pt}{10pt}
	\geometry{
		top=20mm,
		total={155mm,230mm}}
	\newtheorem{theorem}{{\textsc{Theorem}}}[section]

	\newtheorem*{postulate}{\textsc{Postulate}}
	\theoremstyle{definition}

	\newcommand{\suchthat}{\mathrel{\mathop\supset}\kern-4.0pt-\kern-1.0pt-~}
	
	\usepackage{tocloft}
	\cftsetindents{chapter}{0em}{1.9em}
	\cftsetindents{section}{2em}{3em}
	\cftsetindents{subsection}{5em}{3.5em}
	\cftsetindents{subsubsection}{8.5em}{4em}
	
	\setcounter{tocdepth}{3}
	\newcommand{\tocseparator}{%
		\addtocontents{toc}
		{\protect\addvspace{10pt}%
			\hrule
			\protect\addvspace{5pt}%
		}%
	}
	\preto\backmatter{\tocseparator}
	
	\newcommand\bulgelinesmall{\begin{center}	
			\begin{tikzpicture}[x=0.5\textwidth]
				\draw[outer color=black,inner color=gray!60!black] circle[x radius=0.5,y radius=0.014];
			\end{tikzpicture}
	\end{center}}

	\newcommand\rontoc{
		\phantomsection\addcontentsline{toc}{chapter}{{Table of Contents}}
		\tableofcontents\clearpage\pagenumbering{arabic}
	}
	
	\newmdenv[tikzsetting={draw=black, fill=white,},settings={\tikzset{every picture/.style={opacity=0.9}}}]{myenvironment}
	
	\newcommand\rontitleaa[5]{
		\begin{titlepage}
			\thispagestyle{empty}
			\newgeometry{top=1.5cm, bottom=.5cm}
			\begin{mdframed}[rightline=false, leftline=false, topline=false, bottomline=false]
				\rule{0pt}{0.015\textheight}
				\begin{center}
					\vspace{-3em}
					{\large \textsc{\hspace{4pt}#1  \\[0.2\baselineskip] }}		
					
					\vspace{1.5em}
					{{#2}}\\[0.7\baselineskip]
					
					\vspace{0em}
					\bulgelinesmall
					
					{\large \textsc{#3 \\[0.2\baselineskip] }}		
					
					\vspace{2.5em}
					\begin{center}
						\begin{minipage}[t]{1\textwidth}
							\begin{abstract}
								#4
							\end{abstract}
						\end{minipage}
					\end{center}
					
					\makebox[0pt][l]{\rule{0pt}{0.2\textheight}}
					
					\rule{0pt}{0.02\textheight}
				\end{center}
			\end{mdframed}
			\clearpage
		\end{titlepage}
		\pagecolor{white}
	}

	\DeclareMathOperator*{\arginf}{arg\,inf}

\usepackage{extarrows}
\usepackage{setspace}
\usepackage{schemata}
\newcommand\sch[4]{\Schema{#1}{#2}{\schemabox{#3}}{\schemabox{#4}}}
\begin{document}
{\rontitleaa{ }{\Huge\bfseries\textsc{\spaceout[1pt]{Emergence of Time in a \\\onehalfspacing\onehalfspacing\onehalfspacing\onehalfspacing\onehalfspacing\onehalfspacing\onehalfspacing\onehalfspacing Participatory Universe}}}{\color{darkgrey}{Bharath Ron}\\ {\small University of Vienna}}{
		After stating the measurement problem, physicists usually assume the problem to be coming from the measurement part. Since classical probabilities also collapse when updating information, there is nothing special about quantum state collapse. I believe the problem comes from the unitary evolution part of quantum theory. The question we should be asking is not `what happens during measurement?', but `what is time?'. After discussing the problems with time evolution in quantum theory, we propose a new approach to interpret time and argue how it would emerge from the non-commutativity of quantum theory, assuming participatory realism. Its relation to the familiar mechanical/unitary notion of time is discussed.
		
		\hspace{1em} The subjectivity associated with the increasingly popular epistemic interpretations of quantum theory makes it look like it’s a cure that is worse than the disease. We attempt to get objectivity back into quantum physics without resorting to the many world or other such interpretations. \\\par
		\vspace{1em}
		}
	\clearpage
	{\pagenumbering{roman}
		\clearpage
		\rontoc}}
\chapter{\textsc{Statement of the Problem}}

\noindent The path to the construction of a new physical theory is usually a complicated one. In its initial stages, lots of ideas are tried and tested. The surviving ideas are revised, modified, and more clearly defined. These become the postulates of the theory. The purpose of this chapter is however not to study the historical construction of quantum theory but to give instrumentalist reconstruction of the formal structure of quantum theory. The purpose of reconstruction is to translate primary scientific ideas into logically impeccable language such that the new presentation is superior concerning clarity and precision. Reconstruction makes the core underlying structure of physical theory clearer and could expose the problems. Many believe that quantum physics is so different from the usual way of thinking that its meaning cannot be communicated directly. People belonging to this group have a `shut up and calculate' attitude towards the subject and they intend to acquire an intuitive feeling for the subject through calculation and solving problems. We attempt, through reconstruction, to show them that some of the foundations of quantum mechanics can be clarified and the problems can be clearly laid out. We will divide quantum theory into two parts, the semantics part and the evolution part. The semantic part will consist of the physical objects of the theory which in our case are observables and states and their interpretation. The evolution part is about the dynamic processes of the physical theory. This chapter aims to get to a clear formulation of the measurement problem. The main idea of this paper is in the subsection \S\ref{section:Time}, rest of this paper is just carefully implementing the idea.

The presentation here will be closest to G\"unther Ludwig school with some stuff borrowed from quantum logic literature. The Ludwig school has the advantage of being compatible with most interpretations if suitably reformulated. 
\section{Semantics Part I: Physical Objects}
A physical theory is in some sense to be interpreted from outside in terms of pre-theories not belonging to the theory in question itself. To minimize going to these pre-theories we adopt a purely instrumentalist view of physics. The construction and behavior of instruments will not be of interest to us. Any changes occurring in the instruments during `measurements' will be accepted as objective events. According to this point of view, the fundamental notions of quantum mechanics have to be defined operationally in terms of macroscopic instruments and prescriptions for their application. Quantum mechanics is then interpreted entirely in terms of such instruments and events. These instruments and events are our links to `objective reality'.

\subsection{Effects and Ensembles}
From this instrumentalist or operational point of view, the notion of `state' can be defined in terms of the preparation procedure. A preparation procedure is characterized by the kind of system it prepares. The other important thing is the existence of a measuring instrument that is capable of undergoing changes upon their interaction. The observable change in the instrument is called an effect in the Ludwig school.

To simplify the procedure consider instruments that record `hits'. These instruments perform simple `yes-no' measurements. Any measurement can be interpreted as a combination of yes-no measurements. These yes-no instruments can be used to build any general instrument. Suppose we have such an instrument, label its registration procedure by $R$. If the experiment is conducted a lot of times, we get a relative frequency of occurrence of `yes'. To every preparation procedure $\rho$ and registration procedure $R_i$ there exists a probability $\mu(\rho, R_i)$ of occurrence of `yes' associated with the pair.
$$(\rho,R_i)\longrightarrow \mu(\rho|R_i).$$
The numbers $\mu(\rho|R_i)$ are called operational statistics. Two completely different preparation procedures may give the same probabilities for all experiments $R$. Such preparation procedures must be considered equivalent. Such preparation procedures are called operationally equivalent preparations. A precursor to the notion of a state of the system is an equivalence class of preparations procedures yielding the same result. They are called ensembles. 

The basic mathematical structure of ensembles and effects can be understood using purely mathematical reasons, without introducing any new physical law. Denote the class of ensembles by ${S}$ and the class of effects by ${E}$. The maps of interest to us are the following,
$${S}\times {E}\xlongrightarrow{\mu}[0,1].$$
There may be two experiments that give the same probabilities for every ensemble. Such apparatuses must be considered equivalent. They are called operationally equivalent effects. An effect is the equivalence class of apparatuses yielding the same result. In general, a registration procedure $R$ for an experiment will have outcomes $\{R_i\}$. For an outcome, $R_i$ of the registration procedure $R$, denotes the corresponding equivalence class of measurement procedures by $E_{R_i}$. Each outcome $R_i$ of the registration procedure corresponds to a functional $E_{R_i}$ called the effect of $R_i$ that acts on the ensemble of the system to yield the corresponding probability.
$$E_{R_i}:\rho\mapsto E_{R_i}(\rho)=\mu(\rho|R_i).$$
Maps of interest to us will be those that assign to each of its outcomes $R_i$ its associated effect $E_{R_i}$. Since each ensemble fixes a probability distribution we have,
$$\mu_\rho:R_i\mapsto \mu_\rho(R_i)=\mu(\rho|R_i).$$
The above-given map $\mu_\rho$ is determined by the instrument and the registration procedure. Accounting to the fact that preparation procedures can be combined to produce a mixed ensemble, the set of ensembles is taken to be a convex set. Since a mixture of ensembles corresponds to a convex combination of probabilities each functional $E_{R_i}$ preserves the convex structure. Since two preparations giving the same result on every effect represent the same ensemble and two measurement procedures that can't distinguish ensemble represent the same effect, ensembles and effects are mutually separating. A generalized probabilistic theory is an association of a convex state space and effect vectors to a given system, such that the states and effects are uniquely determined by the probabilities they produce. This is known as the principle of tomography. The aim is to obtain a GPT from an operational theory. We are interested in embedding the ensembles inside the vector space of linear functionals on the effects and embed effects inside the vector space of linear functionals on the ensembles. More generally, one takes an operational theory and `quotients' with operational equivalences to obtain a GPT.

Denote by $\mathcal{S}$ the set of maps, $f:E\longrightarrow\mathbb{R}$ such that $f(X)=\sum_i \alpha_i \mu(\rho_i |X)$ and denote by $\mathcal{E}$ the set of maps, $g:S\longrightarrow\mathbb{R}$ such that $g(\rho)=\sum_i\beta_i \mu(\rho | R_i)$ where $\rho_i$ and $R_i$ are ensembles and effects respectively and $\alpha_i ,\beta_i\in\mathbb{R}$. Clearly $\mathcal{S}$ and $\mathcal{E}$ are real vector spaces. We can embed ensembles inside $\mathcal{S}$ with the map, 
$$\rho\longmapsto \mu_\rho,$$
and similarly embed effects inside $\mathcal{E}$ with the map, 
$$R_i\longmapsto E_{R_i}.$$
The bilinear map $\langle\cdot|\cdot\rangle : \mathcal{S}\times \mathcal{E}\to \mathbb{R}$ which coincides with $\mu$ is then uniquely determined. $\langle \mathcal{S}\:|\: \mathcal{E}\rangle$ becomes a dual pair. The completions of $\mathcal{S}$ and $\mathcal{E}$ will provide us the necessary mathematical structure for ensembles and effects. We will denote $\langle \cdot |\cdot \rangle$ by $\mu$. 

A registration procedure $E_R$ is an effect valued function that assigns to each possible outcome $R_i$ its effect $E_{R_i}$, 
$$E_R:\:R_i\longmapsto E_{R_i}.$$
It's important to find a mathematical structure that describes the registration procedure $E_R$ beyond this basic vector space structure. The purpose of this section is to study the mathematical representatives of effects and ensembles in the quantum formalism.

To get the mathematical representatives of physical observables one has to study the logical relations of a set of propositions that are considered meaningful and empirically verifiable according to the theory that describes the physical system. The logic of a physical system will mean the algebraic structure that represents the equivalence classes of the elementary sentences.  To simplify the procedure one initially reduces the elementary sentences of the system to simple `yes-no' questions called propositions. For the development of any mathematical theory, the first step is the idealization of the registrations. Here we are satisfied with the usage of real numbers for the purpose of labeling the outcomes. 

The concept of observable which is one of the main physical objects of quantum theory can be obtained from a certain idealization of the registration procedure. Consider a registration procedure $E_A$ whose outcomes $\{A_i\}$ are measured using the same equipment. The events of such a registration procedure should form a Boolean ring. The aim is to arrive at the notion of observable from these special kinds of registration procedures. What we seek are maps from Boolean rings to the effects. A mapping $A$ of a Boolean ring $\Sigma$ into an ordered interval $[0,\epsilon]$ of a vector space, such that, $A(\mathbb{I})=\epsilon$ where $\mathbb{I}$ is unit of $\Sigma$ and 
$$A(\sigma_1 \vee\sigma_2)=A(\sigma_1)+A(\sigma_2)\:\:\text{ for all } \:\sigma_1\wedge \sigma_2=0,$$
is called an additive measure on $\Sigma$. A set $F\subset \mathcal{E}$ is called a set of coexistent effects if there exists a Boolean ring $\Sigma_A$ with an additive measure $A:\Sigma_A\xrightarrow{}\mathcal{E}$ such that $F\subset A\Sigma_A$. 

An observable is a special kind of registration procedure where the outcomes form a complete Boolean ring. An observable is a pair $(\Sigma_A, A)$, where $\Sigma_A$ is a Boolean ring and $A$ is an additive measure, 
\begin{align*}
	A:\:\Sigma_A\to\mathcal{E}.
\end{align*}
We will denote the observable by the map $A$. The complete Boolean lattice structure of $\Sigma_A$ is the idealization of the registration procedure. Observables are effect-valued functions where outcomes have a Boolean lattice structure. 

Suppose we have two observables $A$ and $B$ and there exists a homomorphism $h$ of the Boolean ring $\Sigma_A$ into the Boolean ring $\Sigma_B$
then intuitively the observable $B$ measures more than $A$ since the measurements of the observable $A$ is contained in the observable $B$. Two observables are equivalent if the homomorphism $h$ is an isomorphism. Two observables $A$ and $B$ are said to coexist if there exists an observable $AB$ and two homomorphisms $h$ and $i$ such that $h: \Sigma_A\to \Sigma_{AB}$ and $i:\Sigma_B\to \Sigma_{AB}$. 
Denote by $\Xi$ the effects that coexist with every other effect. Two observables $A$ and $B$ are mutually complementary if every coexistent effect is in $\Xi$. If two effects $E_{A_i}\in A$ and $E_{B_j}\in B$ are coexistent then at least one of them is in $\Xi$. The existence of such observables is a feature of quantum mechanics that wasn't the case in classical mechanics. We will skip here the notion of preparator and the interested reader should read Ludwig's original work \cite{Ludwig2},\cite{Ludwig3}.

Suppose $A:\Sigma_A\to \mathcal{E}$ is an observable then a state $\mu_\rho$  gives us a map,
$$\mu^A_\rho:\:\Sigma_A\to [0,1],$$ 
such that $\mu^A_\rho(0)=0$, $\mu^A_\rho(E^\perp)=1-\mu^A_\rho(E)$ and whenever $E_i$ are mutually orthogonal,
$$\mu^A_\rho(\vee_i E_i)=\sum_i\mu^A_\rho(E_i).$$
For all practical purposes, we will assume the measurement scale is separable. This assumption gives us all the nice properties needed to do mathematics.

It is important to note that preparation and registration procedures producing the same ensembles and effects are not always equal, in fact, the notion of equality won't even make sense. The transition from preparation and registration procedures to ensembles and effects is a transition from the real world to the abstract mathematical world. It should be noted that it doesn't make sense to `prepare' closed systems, one has to assume such systems start off in some state a priori. 

\subsection{Observables and States}
By the end of the nineteenth century, it was clear that elementary processes obeyed some `discontinuous' laws. There existed no mathematical formalism of quantum theory that would provide a unified structure. Heisenberg's solution to this problem was to use linear operators as a starting point. The space of functions on both discrete and continuous spaces have the same Hilbert space structure. The coexistence of discrete and continuous observables is possible. The necessary structure for the abstract mathematical framework of quantum theory is found in Hilbert spaces and operator algebras.

Let $(\mathcal{H},\langle \cdot|\cdot\rangle)$ be a complex Hilbert space. $\mathcal{P}(\mathcal{H})$ denote the set of all closed subspaces. Denote $\mathcal{H}_i\leq \mathcal{H}_j$ if and only if $\mathcal{H}_i \subseteq \mathcal{H}_j$. The relation $\leq $ is a partial ordering in $\mathcal{P}(\mathcal{H})$. Join $\vee$ of a family $\{\mathcal{H}_i\}_{i\in I}$ is the linear span of the family denoted $\vee_i \mathcal{H}_i$. Meet $\wedge$ of a family $\{\mathcal{H}_i\}_{i\in I}$ is the intersection of the family, denoted $\wedge_i \mathcal{H}_i$.  The orthocomplement of $\mathcal{H}_i$ in $\mathcal{P}(\mathcal{H})$ denoted by $\mathcal{H}_i^\perp$ is the closed subspace of vectors $\varphi\in\mathcal{H}$ such that $\langle\varphi|\mathcal{H}_i\rangle =0$. Since there is a bijection between closed subspaces of a Hilbert space and projection operators acting on the Hilbert space, the set of all projection operators on the Hilbert space inherits a lattice structure from the lattice of closed subspaces. Abusing notation, we will denote the projection operators on $\mathcal{H}$ by $\mathcal{P}(\mathcal{H})$. The orthocomplement of the projection $E$ is the projection onto the orthogonal complement of the subspace corresponding to the projection operator $E$ and is denoted by $E^\perp$. The lattice structure of $\mathcal{P}(\mathcal{H})$ coming from the above relations gives us the necessary structure to get the mathematical representatives of physical observables. The non-Boolean lattice $\mathcal{P}(\mathcal{H})$ of projections should act as the space of effects.
$$\mathcal{E}
\equiv \mathcal{P}(\mathcal{H})$$
For a family of projection operators to represent an observable, we should make sure that the family forms a Boolean algebra. A quantum mechanical observable is an additive measure of the form,
$$E_A:\: \Sigma_A\to \mathcal{P}(\mathcal{H}),$$
a projection valued function. Usually in physical experiments, the statements that can be made are of the type `the value of the observable lies in some set $\epsilon_i$ of real numbers'.  To accommodate the fact that measurement scale is composed of real numbers, we identify $\Sigma_A$ with the Borel sets of $\mathbb{R}$. {It should be noted that the observables need not be real, the physics community has historically decided to use real numbers to label the outcomes of experiments. Any other labeling should work equally well.} D\"oring and Isham have done an interesting generalization of this scheme \cite{doring}. Their idea seems to be to replace the Boolean structure in $\Sigma_A$ with a more general propositional language system and question if values of the system should be more general than `real'. Though we find this to be a beautiful generalization for the future of quantum theory we don't think this is the part needing fixing for solving the foundational problems in quantum theory. We believe we can get a lot of work done with real measurement scales itself. 

The quantum observables are analogous to classical random variables, namely, that of a projection valued measure,
$$E_A: \:\mathcal{B}(\mathbb{R})\to \mathcal{P}(\mathcal{H}).$$
This generalizes the classical case, for which mathematical representatives were the measure space $(\Omega,\Sigma(\Omega),\mu)$, where the $\sigma$-algebra, $\Sigma(\Omega)$ is a class of subsets of the set $\Omega$ which correspond to events and $\mu$ is a probability measure. A classical random variable is defined as a map $X: \Omega\to \mathbb{R}$. The map doing the work in assigning necessary probabilities is its inverse, considered as a set map,
$$X^{-1}: \:\mathcal{B}(\mathbb{R})\to \Sigma(\Omega).$$
A spectral measure is a projection operator-valued function $E$ defined on the sets of $\mathbb{R}$ such that, $E(\mathbb{R})=I$ and $E(\sqcup_i \epsilon_i)=\sum_i E(\epsilon_i)$, where $\epsilon_i$s are disjoint Borel sets of $\mathbb{R}$. The spectral theorem says that every self-adjoint operator $A$ corresponds a spectral measure $E_A$ such that,
$$A=\int \lambda \:dE_A(\lambda),$$
and conversely, every spectral measure corresponds to a self-adjoint operator. In the finite-dimensional case this reduces to $A=\sum_i\lambda_i E_i$ where $E_i$s are projections onto eigenspaces of $\lambda_i$s. Observables in quantum theories are represented by self-adjoint operators on some complex Hilbert space and the orthogonal projections of the self-adjoint operator correspond to the events. The values of the observable are the spectrum of the operator. The characteristic feature of quantum theory is that the space of effects is a non-commutative entity.

The mathematical representatives of the physical states for the quantum case are the maps, $\omega:\mathcal{P}(\mathcal{H})\to [0,1]$, such that $\omega(0)=0$, $\omega(E^\perp)=1-\omega(E)$ and $\omega(\vee_i E_i)=\sum_i \omega(E_i)$ for mutually orthogonal $E_i$. For an observable with the associated self-adjoint operator $A$, the map 
$$\mu^A=\omega\circ  E_A:\:\Sigma_A\to [0,1],$$ 
determines a classical probability measure. The existence and classification of such non-commutative probability measures on Hilbert spaces is given by the Gleason's theorem.

\vspace{.5em}
\begin{theorem}
	{\bfseries{\textsc{(Gleason) }}} If the complex separable Hilbert spaces $\mathcal{H}$ of dimension greater than 2, then every $\omega$ is of the form
	$$\omega(E)=Tr(\rho E).$$
	where $\rho$ is a positive semidefinite self-adjoint operator of unit trace or density matrix. Conversely, every density matrix determines a state as defined in the above formula.\qed
\end{theorem}
\vspace{.5em}

For more general quantum experiments correspond to positive operator-valued measures\footnote{The effects $E$ are given by positive operators, $O\leq E\leq I$ as probabilities are positive quantities. Since these should sum to 1 for an experiment, it will be a resolution of identity $\sum_i E_{A_i}=I$, where $E_{A_i}$s are effects. The resolution of identity $E_A:A_i\to E_{A_i}$ is called positive operator-valued measure (POVM). General quantum mechanical experiments are represented by pairs $(\rho, E_A)$. For Gleason's theorem in this setting see \cite{Busch3}.}. We call $\omega$ a Gleason measure. Every state corresponds to a positive semidefinite self-adjoint operator of unit trace. We denote the set of all states on the Hilbert space $\mathcal{H}$ by $\mathcal{S}(\mathcal{H})$. The mathematical representatives of ensembles are states.
$$\mathcal{S}\equiv \mathcal{S}(\mathcal{H})$$
The extreme points of this convex set are called pure states. For an observable with an associated self-adjoint operator $A$ the probability that the observable takes a value lying in the interval $\epsilon$ is given by, 
$$\mu^A_\rho(\epsilon)=Tr(\rho E_A(\epsilon)).$$
The expectation value of the observable will be,
$$\langle A\rangle =\int \lambda\: d\mu^A_\rho(\lambda)=Tr(\rho A).$$
All the Hilbert spaces will be assumed to be separable, complex. 

Given a finite number of Hilbert spaces $\mathcal{H}_i$, for $n$ quantum systems, the problem is to describe the Hilbert space appropriate to the `product' system.
Let $\mathcal{I}$ denote a possible solution to this problem: that is the states of $\mathcal{I}$ are supposed to be states for the product system. Then, at the very least, some of the preparation procedures for the product system should be obtainable by arranging in some manner the preparation procedures on the individual systems. We should be able to construct a certain function,
$$f:\:\mathcal{H}_1\times\cdots\times \mathcal{H}_n \to \mathcal{I}.$$
The interpretation of $f$ is that, it introduces a component from each individual system into the product system. Accounting to the superpositions, the product system should inherit the structure from the components. The map $f$ must be linear for each component. The universal solution $\mathcal{H}$ to this problem is the algebraic tensor product. It's the vector space $\mathcal{H}$ together with an $n$-linear map $t$ such that, for any $n$-linear map $f:	\mathcal{H}_1\times\cdots\times \mathcal{H}_n\to\mathcal{I}$, there exists a unique linear map $\overline{f}:\mathcal{H}\to\mathcal{I}$,

\[ \begin{tikzcd}
	\mathcal{H}_1\times\cdots\times \mathcal{H}_n \arrow{r}{t} \arrow[swap]{rd}{f} & \mathcal{H} \arrow{d}{\exists !\:\:\overline{f}} \\%
	&\mathcal{I}
\end{tikzcd}
\]
This vector space inherits a canonical inner product from the component Hilbert spaces. The completion of $\mathcal{H}=\otimes_{i\in I}\mathcal{H}_i$ under the canonical inner product will serve as the Hilbert spaces for product systems.

It's important to note that when we are given a closed system, the notion of preparation of state doesn't make sense. So in such cases, we are stuck with states given by nature.

\subsubsection{von Neumann Lattices}
A further abstraction of the mathematical representatives is necessary for quantum theory to be applicable to a wider range of phenomena. All we need is the algebraic structure of the observables and states.\footnote{The only use of this part is in the discussion of thermal time hypothesis later in section \S\ref{section:thermal} and can hence be skipped.} A linear operator $A$ defined on the Hilbert space $\mathcal{H}$ is bounded if, $\|A\|:=\sup_{\|\varphi\|\leq 1}\|A\varphi\|< \infty.$ Denote by $\mathcal{B}(\mathcal{H})$ the bounded operators on the Hilbert space $\mathcal{H}$. $\mathcal{B}(\mathcal{H})$ itself is a vector space over complex field. With respect to the norm $\|\cdot\|$, the vector space $\mathcal{B}(\mathcal{H})$ is a Banach space. The topology determined by the above norm is called uniform topology. The product on $\mathcal{B}(\mathcal{H})$ is defined by the composition. $\mathcal{B}(\mathcal{H})$ is a Banach algebra and it holds that,
$$\|AB\|\leq \|A\|\|B\|.$$
Further, $A^*$ is called the adjoint of $A$ if, $\langle \varphi |A \varkappa\rangle =\langle A^*\varphi|\varkappa\rangle.$ for all $\varphi,\varkappa\in\mathcal{H}$. The map $*: A\mapsto A^*,$ is conjugate linear and satisfies the properties, $(A^*)^*=A$, $(AB)^*=B^*A^*$ and $\|A^*\|=\|A\|$. A Banach algebra with these properties is called an involutive Banach algebra. A $C^*$-algebra is an involutive Banach algebra such that the following holds;
$$\|A^*A\|=\|A\|^2.$$
The algebra of operators that corresponds to observables will be called an algebra of observables. Since the state is characterized by the expectations it defines, one may consider the expectation functionals on $C^*$-algebras as a starting point. What we are interested in are functionals on the algebra of observables $\mathcal{A}$. The expectation should be real. The linear functional when restricted to self-adjoint operators will be real-valued. The expectation of positive operators should be positive. 

For a $C^*$-algebra $\mathcal{A}$, a state is a linear functional, $\omega: \mathcal{A}\to \mathbb{C}.$ Such that $\omega|_{\mathcal{A}_0}:\mathcal{A}_0\to \mathbb{R}$. where $\mathcal{A}_0$ are self-adjoint operators in $\mathcal{A}$ and satisfies $\omega({I})=1$ and
$$\omega(A^*A)\geq 0 \:\:\:\: \text{ for all }A\in\mathcal{A}.$$
The set of all continuous linear functionals over $\mathcal{A}$ is denoted by $\mathcal{A}^\#$. $\mathcal{A}^\#$ is itself a Banach space. The set of states is a convex subset of $\mathcal{A}^\#$ which we will denote by $S(\mathcal{A})$. 

The Gelfand-Naimark-Segal theorem constructs for a given $C^*$-algebra $\mathcal{A}$ and a state $\omega$ a representation of the algebra of observables $\mathcal{A}$ on the set of bounded operators $\mathcal{B}(\mathcal{H})$ for some $\mathcal{H}$. Given a state $\omega$ on $\mathcal{A}$ one considers the two-sided ideal in $\mathcal{A}$ defined by,
$$J_\omega=\{A\:|\:\omega(A^*A)=0\}.$$
Factorizing $\mathcal{A}$ by $J_\omega$ an inner product is introduced on the quotient space $\mathcal{A}_{J_\omega}$ defined by,
$$\langle [A]|[B]\rangle :=\omega(A^*B).$$
where $[A]$ and $[B]$ denote the equivalence classes determined by $A$ and $B$ respectively. The new vector space is completed by adding all the Cauchy sequences and we denote the Hilbert space by $\mathcal{H}_\omega$. On this Hilbert space we have the representation of the algebra $\mathcal{A}$, $\pi_\omega:\mathcal{A}\to \mathcal{B}(\mathcal{H}_\omega)$ defined by,
$$\pi_\omega(A)[B]\equiv[AB].$$
Let $[{I}]=\Omega_\omega$. The expectation of any observable can then be written as,
$$\omega(A)=\langle \Omega_\omega|A\Omega_\omega\rangle.$$
A state on the algebra can be represented as a vector in some Hilbert space.
A vector $\varphi\in\mathcal{H}$ is said to be cyclic for $\mathcal{A}$ if the closure of $\mathcal{A}\varphi$ is same as $\mathcal{H}$. A vector $\varphi$ is separating for $\mathcal{A}$ if $A\varphi=0$ implies $A=0$ for all $A\in\mathcal{A}$. The vector $\Omega_\omega$ is cyclic for $\mathcal{A}$.

\noindent The topology determined by the seminorms $$p_{\varphi,\varkappa}(A)=\langle \varphi | A \varkappa\rangle,$$ on the algebra $\mathcal{A}$ is called weak topology. It's the smallest topology with respect to which the maps $p_{\varphi,\varkappa}$ are continuous. So $A_n\to A$ if $p_{\varphi,\varkappa}(A_n)\to p_{\varphi,\varkappa}(A)$. Basis for the topology consists of sets of the form, $W(X,\varphi,\varkappa,\epsilon)=\{Y\:|\:\langle \varphi|(Y-X) \varkappa\rangle <\epsilon\}$. The expectations give us seminorms on the space of operators and the topology determined by them is called ultraweak topology. Ultraweak topology is the smallest topology with respect to which the expectation functionals
$$p_\rho(A)=Tr(\rho A).$$
are continuous. Intuitively we expect the expectations of observables close to each other to be close. The weak topology and ultraweak topology capture this notion of convergence. von Neumann algebras are those algebras that are closed under the ultraweak topology. The objects of interest to us in the general quantum case are von Neumann algebras.

It's important to note that we don't lose anything by considering only the bounded operators. The adjoint of an operator $A$ is defined as the operator $A^\dagger$ such that $\langle \varphi|A\varkappa\rangle=\langle A^\dagger \varphi|\varkappa\rangle$ for $\varphi,\varkappa\:\in\mathcal{H}$. If the operator is unbounded there will exist vectors $\vartheta$ for which $\|A\vartheta\|\to\infty$ and we can't define the adjoint on the whole Hilbert space. In such cases, one has to be careful about the domain of the operator to define the adjoint. It's also not physically meaningless to consider only the bounded operators on the Hilbert space. If $A$ is an unbounded observable the physical reasoning could be that experiments can take only an arbitrarily large but finite range of values $[-n,n]$. If the spectral decomposition of the unbounded operator is,
$$A=\int \lambda \:dE_A(\lambda),$$
We have by restricting to the interval $[-n,n]$ a bounded operator, $A_n=\int_{\lambda\in[-n,n]} \lambda \:dE_A(\lambda)$. Since $A_n\to A$, and Hilbert spaces are complete, the operators $\{A_n\}$ contain all the physical information contained in the operator $A$.

The commutant of an algebra $\mathcal{A}\subseteq \mathcal{B}(\mathcal{H})$ is the set of operators in $\mathcal{B}(\mathcal{H})$ that commute with every element of $\mathcal{A}$. 
$$\mathcal{A}'\equiv\{B\in\mathcal{B}(\mathcal{H})\:\:|\:\: AB=BA, A\in\mathcal{A}\}.$$
For any set $\mathcal{A}\subset\mathcal{B}(\mathcal{H})$, the commutant $\mathcal{A}'$ will be a subalgebra of  $\mathcal{B}(\mathcal{H})$ and it will be closed with respect to weak, strong and uniform topology on account of multiplication being continuous. If $\mathcal{A}\subset T$ then $T'\subset \mathcal{A}'$. von Neumann double commutant theorem says that $\mathcal{A}\subset \mathcal{B}(\mathcal{H})$ is a von Neumann algebra if and only if $\mathcal{A}=(\mathcal{A}')'$. 

An immediate consequence is that if $A$ is a self-adjoint element in the von Neumann algebra $\mathcal{A}$, then every spectral projection of $A$ belongs to $\mathcal{A}$. Consider the projection operators of the von Neumann algebra $\mathcal{A}$, denoted by $\mathcal{P}(\mathcal{A})\subseteq \mathcal{A}$. $\mathcal{P}(\mathcal{A})'$ is a von Neumann algebra and so is $(\mathcal{P}(\mathcal{A})')'$. Suppose $Q\in \mathcal{P}(\mathcal{A})'$, then the spectral projections of $Q$ also are in $\mathcal{P}(\mathcal{A})'$. Hence the spectral projection of $Q$ commutes with the spectral projection of every self-adjoint operator in $\mathcal{A}$ since they lie in $\mathcal{P}(\mathcal{A})$. $Q$ commutes with every element of $\mathcal{A}$. Hence we have $\mathcal{P}(\mathcal{A})'\subseteq \mathcal{A}'$. It follows that, $\mathcal{P}(\mathcal{A})'=\mathcal{A}'$ and hence,
$$(\mathcal{P}(\mathcal{A})')'=(\mathcal{A}')'=\mathcal{A}.$$
The projection operators in $\mathcal{A}$ generates the algebra $\mathcal{A}$. These projection operators also have a complete lattice structure. The effects in general quantum systems are identified with these von Neumann lattices of projections.


\section{Evolution in Quantum Theories}
Now that the objects of quantum theory are fixed the next step is to study the evolution of the systems. Physical laws axiomatize the regularities found in the evolution of the system. These physical laws act as constraints on the evolution of physical systems. There was a time when one tried to base a physical theory on experimental facts. The physical laws were guessed. When it comes to verification of the laws it can only be stated that all the experiments conducted until then were, within the scope of the inaccuracies, not in contradiction with the law. But how do we come up with these laws? Is it possible to develop a method of guessing physical laws? and why do we trust physical laws? It is sometimes not difficult to guess from experimental data a mathematical representation of this relation. Everything is permitted in the guessing of physical laws. One commonly employed way of finding the physical laws is to take the laws from physical theories with a common application domain and reformulate it in terms of the objects of the new physical theory. Symmetries of the system for example act as common laws for many physical theories. 
It can also be that we trust a new law even if we have made no critical experiments. Before Einstein's new gravitational theory there were two other well-tried theories: relativity and Newton's gravitational theory for different domains of application. But it was not possible to unify these two theories. Einstein succeeded in achieving this by introducing a new law with an enlarged application domain. 

Physics has always been about developing a theoretical framework that explains experimental observations. The experimental observations are roughly speaking `{interpreted}' in terms of this theory. Quantum mechanics is a theoretical framework using which we intend to interpret experiments. Each physical theory consists of physical objects, and the laws of the physical theory provide morphisms between them. A physical theory $PT$ is composed of an application domain $A$, a mathematical theory $MT$, and the correspondence rules $\leftrightarrow$. The correspondence rules $\leftrightarrow$ are prescriptions of how to translate facts detected in the world into mathematical theory. A physical theory is then,
$$PT\equiv A\longleftrightarrow MT.$$
The facts about the physical domain are then `interpreted' using the mathematical theory.  For the description of the facts in the physical domain, one uses `pre-theories' which already come with their own interpretation. Ludwig's instrumentalist approach greatly simplifies this part. The construction and behavior of instruments will not be of interest to us.

Suppose a system prepared in a state $\rho$ undergoes a process without getting destroyed. Assign to this system a quantum state. The original preparation procedure along with the process can be considered a new preparation procedure. The equivalence class of the new preparation procedure will define the new quantum state after the process. This state depends on the original preparation procedure and the process. Each process corresponds to a linear map,
$$\alpha:\mathcal{A}\to\mathcal{B}.$$
The algebras of observables $\mathcal{A}$ and $\mathcal{B}$ represent the input and output systems respectively. To an initial state $\rho$ of $\mathcal{A}$ the channel associates the output state $\rho\circ \alpha$ of $\mathcal{B}$. 

If density matrices are used to describe quantum states in quantum mechanics, then a process must be some operation that sends density matrices to density matrices. So for finite-dimensional state spaces, a process should be a linear map of vector spaces of matrices. It preserves the trace of matrices and takes hermitian matrices with non-negative eigenvalues to hermitian matrices with non-negative eigenvalues. It must take positive operators to positive operators. A map is called positive if it takes positive operators to positive operators.
$$\alpha(P)\geq 0 \:\:\text{ whenever } P\geq 0.$$
Suppose the process acts only on some part of the system then it must still be a process on the total system. The map corresponding to a process should be positive for the bigger system as well. $\alpha\otimes \mathbb{I} : \mathcal{A}\otimes \mathcal{B}(\mathbb{C}^n)\to \mathcal{B}\otimes \mathcal{B}(\mathbb{C}^n)$ is positive. Such maps are called completely positive. A general quantum process corresponds to a completely positive unital mapping.


\subsection{Quantum Measurement}\label{section:measurement}
In quantum theory, the description of the system requires two physical objects. The first being the state of the system which contains the information known about the system. Second, the observables, which are objects about which the information is about. Bayes' theorem says that additional information about a system will alter the probabilities of possible outcomes. The notion of information is closely related to the notion of probability. Probability gives one way to describe information about the events. We are interested in quantifying the amount of information contained in a state relative to another state. 

The relative entropy of two states $\rho$ and $\sigma$ is the informational divergence of $\rho$ from $\sigma$. Suppose the state $\sigma$ contains information only about a subsystem $\mathcal{B}$ of $\mathcal{A}$ and $E$ is a projection of norm one of $\mathcal{A}$ onto $\mathcal{B}$ then the state $\sigma$ should satisfy $\sigma\circ E= \sigma$.  In such a case the informational divergence should have two components. First component is the divergence of $\rho$ from $\sigma$ on the subalgebra $\mathcal{B}$ which is the divergence between the states $\rho|_\mathcal{B}$ and $ \sigma|_\mathcal{B}$. The other component is the remaining information $\rho$ has and this will be the divergence between the states $\rho$ and $ \rho\circ E$. If $R(\cdot\:,\cdot)$ is such a function then,
$$R(\rho, \sigma)= R(\rho|_\mathcal{B}, \sigma|_\mathcal{B})+R(\rho, \rho\circ E).$$
Any automorphism $\alpha$ of the algebra $\mathcal{A}$ should change the information contained in the two states similarly hence the information divergence should be invariant under automorphisms of the algebra,
$$R(\rho, \sigma)= R(\rho\circ \alpha, \sigma\circ \alpha).$$
The informational divergence of a state with respect to itself should be zero $R(\rho, \rho)=0.$ If $R(\cdot\:,\cdot)$ is a real-valued functional satisfying the above conditions then there exists a constant $c\in \mathbb{R}$ such that,
$$R(\rho, \sigma)=c\: \text{Tr}\left( \rho\:(\log \rho-\log \sigma)\right).$$
The relative entropy of the state $\rho$ with respect to $\sigma$ is defined as,
$$J(\rho,\sigma)=\text{Tr}\left( \rho\:(\log \rho-\log \sigma)\right).$$
In the classical case, the Bayes' rule has been shown to be a special case of the constrained maximization of relative entropy \cite{Williams}. The quantum version of this result is obtained in \cite{Kostecki1}. We will state the result here. 

Suppose an observable $A$ has been subjected to measurement. For simplicity we consider the observable to be a discrete observable. Let $A$ be a discrete observable with effects given by the set $\{A_i\}_{i\in I}$ and the corresponding projection operators $\{E_{A_i}\}_{i\in I}$. If the quantum state of the system after the measurement is $\sigma$, it carries information that has to be compatible with the possibility of measuring all eigenvalues of $A$ precisely. Such a situation is given by the condition $[\sigma, A] = 0$. Suppose the result of the measurement is $A_k$ then the probability of measuring $A_k$ again should be $Tr(E_{A_k}\sigma)=1$. Repeated measurements add no new information. The set of all such states such that $Tr(E_{A_k}\sigma)=1$ is a convex set. Let $p=\{p_i\}_{i\in I}$ such that $\sum_ip_i=1$. The set, 
$$\mathcal{S}_p=\{\sigma\in \mathcal{S}(\mathcal{H})\:\:|\:\:[E_{A_i},\sigma]=0,\:Tr(\sigma E_{A_i})=p_i\},$$
encodes the data that the measurement outcome $A_i$ corresponding to the projection $E_{A_i}$ occurs with probability $p_i$. The commutation condition says that they posses a common eigenbasis and also means that $[\sigma, A]=0$.

\begin{theorem}
	{\bfseries{\textsc{(Hellmann-Kami\'nski-Kostecki)}}}
	$$\arginf_{\sigma\in \mathcal{S}_p}\{J(\rho,\sigma)\}=\sum_{i}p_i {E_{A_i}\rho E_{A_i}}/{Tr(E_{A_i}\rho E_{A_i})}.\hfill \qed$$
\end{theorem}

The strong collapse or the L\"uder-von Neumann rule of collapse is a limiting case of the above projection with all $p_i$ going to zero except one. By taking the limit $p_i\to 0$ for $i\neq j$ we get the L\"uder-von Neumann's rule of collapse,
$$\rho\to {{E_{A_j}\rho E_{A_j}}}/{{Tr(E_{A_j}\rho E_{A_j})}}.$$
This amounts to selecting the quantum state that is least distinguishable from the original state among all the states that satisfy the constraint. For proof and generalization of the result to the algebraic case, the interested reader should read the original papers \cite{Kostecki1},\cite{Kostecki3},\cite{Herbut} and the references therein. The origins of principles of maximum entropy and minimum relative entropy go back to the works by Jaynes, Bayer \& Ochs, Ingarden \& Urbanik, Shore \& Johnson, and others \cite{BayerOchs},\cite{Jaynes1},\cite{Ingarden3},\cite{JohnsonShore}. These maps are called measurement channels. In general measurement channels are given by positive operator-valued measures, where for a measure space, $(\Omega,\Sigma)$, and $\epsilon\in\Sigma$ $E(\epsilon)$ is a positive operator, $E(\Omega)=1$ and for pairwise disjoint $\epsilon_i$,
$\sum_i E(\epsilon_i)=E(\vee_i \epsilon_i).$
It should, however, be noted that the L\"uder-von Neumann rule is about calibrating with the experimental result and has no predictivity. We will abuse the notation and denote an event characterized by the effect $E_{A_i}$ by $E_{A_i}$ only.

Since classical probabilities also collapse when updating information, there is nothing special about quantum state collapse.  We will denote the physical objects of quantum theory along with this measurement process by $\mathcal{QM}_M$.

\subsection{Why Unitary?}
Here we give a brief review of the unitary evolution. The purpose of this subsection is to remind ourselves why unitary evolution is used in quantum theory. When it comes to time evolution, the quantum theory continues on with the received view. The symmetries of classical theories are implemented on objects of quantum theory. 

The simplest structure a symmetric map should preserve is the convexity of the space of states, physically corresponding to the fact that a state arises from mixing states with certain statistical weights. Symmetry operations may modify the constituent states but do not change the weights. A bijection $\alpha:\mathcal{S}(\mathcal{H})\to \mathcal{S}(\mathcal{H})$ is a symmetry if it preserves the convex structure of $\mathcal{S}(\mathcal{H})$. For $p_i\in[0,1]$ and $\sum_i p_i=1$,
$$\alpha(\sum_i p_i \rho_i)=\sum_i p_i \alpha(\rho_i).$$
Such a map is called a Kadison automorphism. If a map $\alpha$ is a Kadison automorphism, then Kadison-Wigner theorem says $\alpha$ is of the form,
$$\alpha(\rho)=U\rho \:U^{-1},$$
where $U$ is unitary or antiunitary and is determined up to phase. A unitary operator is a map $U$ such that $\langle Ux, Uy \rangle = \langle x, y \rangle$ and an antiunitary operator is a map $U$ such that $\langle Ux, Uy \rangle = \overline{\langle x, y \rangle}$ where $\langle\cdot,\cdot\rangle$ is the inner product on the Hilbert space.

To implement the symmetries of the system, the symmetries must be represented in terms of Kadison automorphisms. We seek maps from some group to the set of Kadison automorphisms. Whether a specific transformation is unitary or antiunitary depends on its physical nature. Transformations that belong to a continuous group, such as translations and rotations, can only be unitary because in that case any finite transformation can be generated by a sequence of infinitesimal steps. Let $\mathcal{G}$ be the group of symmetries of the system, then to each $g\in\mathcal{G}$ there should correspond a Kadison automorphism,
$$\alpha:\:g\mapsto \alpha_g.$$
By Kadison-Wigner theorem we get a unitary or antiunitary representative $U_g$ to each element $g\in\mathcal{G}$. For now we will assume $U_g$ to be unitary. Given $g,h\in \mathcal{G}$ we know that,
$$\alpha_g\alpha_h=\alpha_{gh}.$$
For compatible representative $U$ we have, 
$$ U_gU_h=\lambda (g,h)U_{gh},$$
where $\lambda (g,h)$ is a phase factor. A map $U:g \mapsto U_g$ satisfying the above relation is called a projective unitary representation. $\lambda (g,h)$s are called multipliers. For $g=e$ we get,
$$U_e=\lambda(e,e)I.$$
We get some conditions on the multipliers $\lambda (g,h)$. Applying several times to $f,g,h$ we get,
$$\lambda (f,g)\lambda (fg,h)=\lambda (g,h)\lambda (f,gh).$$
We also get,
$$\lambda(e,g)=\lambda(g,e)=\lambda(e,e).$$
A projective unitary representation with $\lambda(e,g)=\lambda(g,e)=\lambda(e,e)=1$ for every $g\in\mathcal{G}$ is said to be normalized. A map $g\mapsto U_g$ is called a unitary representation of $\mathcal{G}$ on $\mathcal{H}$ if $U_e=I$ and satisfies,
$$ U_gU_h=U_{gh}.$$
Unitary representations are usually much easier to work with. A theorem of Bargmann says for some groups with nicer properties (connected and simply connected) it's possible to get a unitary representation. One can always consider the universal covering group and get a unitary representation of that anyway.

Given a self-adjoint operator $A$, one can construct a family of unitary operators, $U(t)=e^{-itA}$. Stone's theorem says the opposite is also true. If $t\mapsto U_t$ is a strongly continuous one-parameter unitary group in the complex Hilbert space $\mathcal{H}$, there exists a unique self-adjoint operator $A$ called the generator of the group such that, 
$$U_t=e^{-itA}.$$
We can, therefore, by Stone's theorem, associate with every one-parameter subgroup of $\mathcal{G}$ a unique self-adjoint operator $A_i$. The Lie algebra of the group $\mathcal{G}$ is represented by the self-adjoint operators $A_i$. From the Lie algebra of the group of symmetries, we can obtain the unitary representatives with a factor of $-i$. If the Lie algebra has the basic structure equation, $[a_i,a_j]=\sum c_{ij}^m a_m,$ then the self-adjoint operators $A_i$ corresponding to $a_i$ satisfy the commutator relations, $i[A_i,A_j]=\sum c_{ij}^m A_m.$

We get the Schr\"odinger equation by implementing Galilean symmetries. When the symmetries are taken to be the Galilean group, the time evolution is generated by the Hamiltonian of the system and corresponds to the time translation symmetry of the system.
$$\rho\:\mapsto e^{-itH}\rho \:e^{itH}.$$
This notion of evolution is a direct copy-paste of the classical laws formulated for quantum objects. We will denote the physical objects of quantum theory along with this notion of evolution by $\mathcal{QM}_U$.

\section{The Contradiction}
A physical theory is said to be universal if its domain of application is everything. Thermodynamics for example is considered universal. If an appropriate correspondence from the physical world to objects of thermodynamics exists, the laws of thermodynamics explain the observed phenomena. Classical physics was supposed to be another such universal theory but the experiments of the twentieth century showed it to be not the case. Quantum theory was created as a replacement. Quantum mechanics is supposed to be universal. It is supposed to explain all the observed phenomenon and all non-quantum theories should be approximation theories relative to quantum mechanics. Although the successes of quantum theory may make the idea of the universality of quantum mechanics more compelling, it is important to note that these successes are not proof of the validity of the fundamental principles of the theory. Even if the adherents of the universality of quantum mechanics avoid the problem of elaborating the limits of quantum mechanics, they necessarily introduce a new difficulty, namely, the need to answer the following question: How do we obtain a `determination' of measurement results? This is known as the measurement problem. Here again, we find a variety of different approaches, which range from attempts to show that probability theory itself gives the valid determination to the introduction of consciousness of the observer or to the so-called `many-worlds' interpretation of quantum mechanics. 

What we have are two theories for the same domain of facts, one given by $\mathcal{QM}_M$ and another by $\mathcal{QM}_U$. From the point of view of $\mathcal{QM}_U$, the representation of the Galileo group (or some other group depending on the situation) determines the laws. We shall only examine the condition for the physically important time translation. From this point of view, a description means that for every system there is a corresponding trajectory $\rho(t)$ of the state. What we expect is a reasonable relation between $\mathcal{QM}_M$ and $\mathcal{QM}_U$. There must exist in some sense, an equivalence $N$ between $\mathcal{QM}_M$ and $\mathcal{QM}_U$. 
$$\mathcal{QM}_M\xlongleftrightarrow{N?}\mathcal{QM}_U.$$
According to $\mathcal{QM}_U$, there exists, a measurement of the state `at time' $t$. In $\mathcal{QM}_M$ such a measurement `at time' $t$ is not defined. One can then ask if there exists a theory whose laws reduce to these in special situations. But then there are complications like which law should apply when and why? The decoherence approaches try to `explain' measurement using unitary evolution\footnote{Interested reader should check out \cite{Adler} to understand why decoherence doesn't say jack shit about the measurement problem.} and the stochastic interpretation tries to `explain' unitary evolution through the measurement process. However, we feel neither of them are satisfactory. Others feel there is no need for a unification and that quantum theory in its current form is perfectly fine. People belonging to this group(mostly epistemic interpretation people) view both unitary evolution and measurement process to be perfectly fine and that one should decide whether the unitary evolution applies or measurement process applies depending on the situation. In this case we have two rules for updating the states of the system and it rests on the observer to decide which to apply when. We are however not convinced by this. We seek a universal theory with a single physical idea for the evolution of systems. Such a general theory that contains both doesn't exist in our opinion. The question we should be asking now is if it's possible to find some workaround for this contradiction. 

What we are dealing with are different physical theories. A false view of inter-theory relations has been the source of many false opinions concerning the truth of a physical theory. Thus the misconception has arisen that no theory is really `true' but that during the development of physics a later theory becoming valid makes an older theory untrue. Each physical theory will have its own application domain. What we expect from a better theory is a larger application domain. We can think of physical theories as a category with objects being the physical theories. The inter-theory relations are the morphisms. What we expect is a physical theory from which we can obtain other theories either by approximation or some domain change. This would be a `universal theory'. Quantum theory in its current stage is not a universal theory. Its application domain is not good enough. For example, it cannot explain a phenomenon like gravity. What we seek is a modification of the laws of quantum mechanics that would make it universal. Based on the needs, physical theories are revised to suit us. Once in a while, the change required will be so radical that the objects of the theory themselves need to be replaced. In this case, we are fine with the objects of quantum theory. What we are interested in is a theory consisting of objects of quantum theory and a new law of evolution compatible with measurement that would approximate to $\mathcal{QM}_U$ under the suitable domain of application conditions. We will denote this by $\mathcal{QM}$. 

Physical theories do not begin their development based on some well-defined foundations. The methods of the new physical theory are initially intuitively conceived and applied. The theories usually encounter contradictions on the way, and the discovery of the cause of these contradictions lets us rectify or clarify them. The clarifications for the contradictions help us avoid the contradictions in the future. The contradictions are crucial to the development of any physical theory. It is these clarifications that develop the conceptual foundation of the said physical theory. Einstein's theory of gravity used the old classical physical objects and introduced a new physical law. In quantum mechanics, the physical objects are newly introduced but the physical laws are old received view. We believe this is where the problem comes from. The question we have to ask is not `what happens during measurement?' but `what is time?'. We feel the problem is coming from the unitary evolution part. Synthesis of quantum theory and gravity would require revision of the physical laws in quantum theory. Now the problem is how to guess the new law? As we stated before, one commonly employed way of finding the physical laws is to take the laws from existing physical theories with common application domain and reformulate it in terms of the objects of the new physical theory or try to guess a law that fits with the law of the existing physical theory. One physical theory that has the universality we expect from quantum theory is thermodynamics and we could seek guidance from thermodynamics. 
\newpage
\section{Semantics Part II: Interpretation}
The complications in the interpretation of quantum theory come from the interpretation of probabilities occurring in quantum theory. We follow Adan Cabello's classification of the various interpretations of quantum mechanics \cite{Adan}. Interpretations of quantum theory can be classified according to whether they view probabilities of measurement outcomes as determined or not by intrinsic properties of the observed system. 

\vspace{1.5em}
\hspace{2em}\DoBrackets\sch{-4ex}{5.5ex}{Interpretations}{	
	\DoBrackets\sch{-2.1ex}{2.6ex}{Intrinsic Realism}{
		\DoBrackets\sch{-1.3ex}{1.3ex}{$\psi$-ontic}{Many worlds, etc.\\}
		\\
		\DoBrackets\sch{-1.5ex}{1.7ex}{$\psi$-epistemic}{Ballentine \cite{Ballentine2}\\Spekkens \cite{Spekkens}\\}
	}\\
	\DoBrackets\sch{-2.9ex}{3.8ex}{Participatory Realism}{
		\DoBrackets\sch{-1.3ex}{1.3ex}{About Belief}{
			QBism \cite{Fuchs},\cite{Fuchs2},\cite{FuchsMerminSchack}\\
		}\\
		\DoBrackets\sch{-.1ex}{3.4ex}{About Knowledge}{
			Neo-Copenhagen\\Wheeler \cite{Wheeler},\cite{Wheeler2}\\Relational \cite{RovelliInter}\\Fuchs \& Peres \cite{FuchsPeres}}
	}
}
\vspace{1em}

\subsection{Intrinsic Realism}
The intrinsic realist interpretations are those in which probabilities of measurement outcomes are determined by the intrinsic properties of the observed system. The notion of reality is closely linked with the values of observables in these interpretations. The intrinsic realist interpretation is $\psi$-ontic if they view the quantum state as an intrinsic property of the system. We will only briefly discuss some theorems that discourage us from intrinsic realism and move on. Many world interpretation will not be discussed. A commonsense belief is that at any given moment any physical quantity must have a value even if we do not know what it is. This view comes from our belief in classical physics. This is not problematic in classical physics since the underlying structure for observables and states fits this view perfectly.
\subsubsection{Ontological Model}
In classical physics, the states are specified as points in phase space which correspond to the values of position and momentum, the observables of the system. Such states specify the values of every observable that the system can take. The observer can know with certainty the values of all observables. The states in classical mechanics represent a state of reality. If $\Omega$ is the state space in classical physics, an observable is a map,
$$X:\:\Omega\to \mathbb{R}.$$
Each state would fix a value for the observables. If we denote the valuation map associated with a state $\rho$ by $\lambda_\rho$ then,
\begin{align*}
	\lambda_\rho:\: X\mapsto \lambda_\rho(X).
\end{align*}
All physical quantities possess a value in any state. If $h:\mathbb{R}\to\mathbb{R}$ is a real-valued function we can construct new observables from old ones, the values of which are $h(X):=h\circ X:\Omega\to\mathbb{R}$. In such cases, we should expect the valuation of the new observable to be,
$$\lambda_\rho(h(X))=h(\lambda_\rho(X)).$$
The observable $h(X)$ is defined by saying that its value in any state is the result of applying the function $h$ to the value of $X$.

Effects in the quantum case are projection operators. If a valuation as above exists then the valuation map associated with a state of reality should assign to each projection operator the values $1$ or $0$ based on whether the system was measured with the said property or not. Such maps are called valuation maps or valuations. If $\lambda$ is such a valuation map then $\lambda(\mathbb{I})=1$. If $A$ and $B$ are self-adjoint operators such that for some real-valued function $h$, $B=h(A)$ then, 
$$\lambda(B)=h(\lambda(A)).$$
Hence a valuation map associated with a state $\rho$ is a homomorphism from the algebra of projection operators to the set $\{0,1\}$.
$$\lambda_\rho:\mathcal{P}(\mathcal{H})\to \{0,1\}.$$
Assuming such a valuation map exists, it must satisfy the valuation conditions $\lambda_\rho(\mathbb{I})=1$ and $\lambda_\rho(\sum_i E_i)=\sum_i \lambda_\rho(E_i).$
Such a map satisfies the conditions of Gleason's theorem, hence must take continuous values in $[0,1]$. Since valuation maps can only take discrete values $\{0,1\}$, such a map cannot exist.

\vspace{.5em}
\begin{theorem}
	{\bfseries{\textsc{(Kochen-Specker) }}} If  $\dim (\mathcal{H}) \geq 3$ then there exist no valuations. \qed
\end{theorem}
\vspace{.5em}

The theorem roughly asserts that it is impossible to assign values to all physical observables while simultaneously preserving the functional relations between them. It should, however, be noted that when restricted to commutative subalgebras valuations do exist. Due to the non-commutativity of quantum theory, the values of all the observables can't be known at once and any such notion has to be contextual. The states in quantum theories cannot be interpreted completely ontically. Non-contextual hidden variable theories are also not viable. A `state of reality' is meaningless in quantum theory. Any attempt to view the quantum state as ontic states would require serious mutilation of objects of quantum theory.

The interpretation is $\psi$-epistemic if they view the quantum state as containing knowledge about an underlying reality similar to how we view states in classical statistical mechanics. The point of view given in \cite{Ballentine2} is the statistical interpretation, the quantum states represent partial knowledge about an underlying state of reality.  In classical statistical mechanics, probability distributions are introduced on the phase space. These distributions represent the likelihood of the occurrence of the values. However, if the position and momenta of all the particles are known then we have complete knowledge of the system. These states of complete knowledge of the system correspond to Dirac delta distributions which are in a one-to-one correspondence with the points in phase space. The PBR theorem \cite{Pusey} is a contradiction to interpreting quantum states statistically. 

A non-contextual ontological model of an operational theory is an attempt to provide a causal explanation of the operational statistics. It says that the response of the measurement is determined by the ontic state $\lambda$ of the system, while preparation procedures determine the distribution over the space of ontic states, $\Lambda$, from which $\lambda$ is sampled. An ontological model associates to each preparation $\rho$ a probability distribution $\mu_\rho$ representing the agents' knowledge of the ontic state given the preparation $\rho$. If we denote the set of such distributions by $\mathcal{D}(\lambda)$, the ontological model specifies a map $\mu:S\to\mathcal{D}(\Lambda)$. An ontological model associates to each operational effect a response function on $\Lambda$ representing the probability assigned to the outcome $R_i$ in a measurement of $R$ if the ontic state of the system fed into the measurement device were known to be $\lambda\in \Lambda$. If we denote the set of response functions by $\mathcal{F}(\Lambda)$, the ontological model specifies a map, $\eta:E\to \mathcal{F}(\Lambda)$. These two maps must preserve the convex structure i.e, if $\rho$ is a mixture of $\rho_1$ and $\rho_2$ with weights $\lambda$ and $1-\lambda$ then $\mu_\rho=\lambda \mu_{\rho_1}+(1-\lambda)\mu_{\rho_2}$ and similarly for effects. Furthermore, an ontological model should produce the same probability rule as the operational theory. Assuming $\Lambda $ is discrete for simplicity, we have,
$$\mu(\rho, R_i)=\sum_\lambda \eta_{R_i}(\lambda)\mu_\rho(\lambda).$$
An ontological model of an operational theory is said to satisfy the generalized noncontextuality if every two operationally equivalent procedures have identical representations in the ontological model. That is to say, $\rho\sim \rho' \implies \mu_\rho=\mu_{\rho'}$ and similarly for effects. The same can be defined for GPTs and it's shown in \cite{Schmid} that they are equivalent. We will now go back to the quantum case.

Suppose quantum state $\rho$ is a state of knowledge, representing the uncertainty about the real underlying ontic state of the system $\lambda$. The quantum state $\rho$ results in a physical state $\lambda$ with a probability distribution $\mu_\rho(\lambda)$. If the distributions for distinct quantum states do not overlap then the quantum state can be uniquely inferred from the physical state. If the distributions overlap, then the quantum states can be said to only contain some knowledge about the physical state. Suppose we have two quantum states $\rho_1 $ and $\rho_2$ with overlapping distributions, $\mu_{\rho_1}(\lambda)$ and $\mu_{\rho_2}(\lambda)$ then for any $\lambda$ in the overlap $\Delta$, there is a $q>0$ probability that the physical state is compatible with both quantum states. Now consider two uncorrelated systems that are prepared with two copies of the same preparation device. If the physical states $\lambda_1$ and $\lambda_2$ lie in the overlap $\Delta$, there must be some $q>0$ such that with $q^2$ probability the quantum states $\rho_1\otimes \rho_1$, $\rho_1\otimes \rho_2$, $\rho_2\otimes \rho_1$ and $\rho_2\otimes \rho_2$ will be in this physical state. The measurement on this system can be cleverly chosen such that the first outcome is orthogonal to the first state, the second outcome orthogonal to the second state, and so on.

To arrive at a contradiction consider $\rho_1=|0\rangle$ and $\rho_2=|+\rangle=(|0\rangle+|1\rangle)/\sqrt 2$, and choose the measurement which projects onto the following orthogonal vectors,
$(|0\rangle\otimes |1\rangle+|1\rangle\otimes|0\rangle)/\sqrt 2$, $(|0\rangle\otimes|-\rangle+|-\rangle\otimes|0\rangle)/\sqrt 2$, $(|+\rangle\otimes|1\rangle+|1\rangle\otimes|+\rangle)/\sqrt 2$ and $(|+\rangle\otimes|-\rangle+|-\rangle\otimes|+\rangle)/\sqrt 2$, where $|-\rangle=(|0\rangle-|1\rangle)/\sqrt 2$.
The measuring device should have been uncertain at least $q^2$ of the time about which quantum state was used in the measurement. 

\vspace{.5em}
\begin{theorem}
	{\bfseries{\textsc{(Pusey-Barret-Rudolph)}}} Quantum state interpreted as information about an objective physical state cannot reproduce the predictions of quantum theory.\qed
\end{theorem}
\vspace{.5em}

This will imply it will give an outcome that is predicted to not happen quantum mechanically. Hence, interpreting quantum states as having information about an underlying objective physical state cannot reproduce the predictions of quantum theory. Kochen-Specker theorem rejects non-contextual hidden variable theories and the PBR theorem rejects the existence of non-contextual ontological model or the statistical interpretation. Frequentist interpretation of probabilities occurring in quantum mechanics is problematic. Kochen-Spekker theorem can be evaded by arguments of the sort `the values are never accurately known'. But we don't like to go down this path. We respect the idealization procedure of quantum theory. 

\subsection{Participatory Realism}
The participatory realist interpretations have seen increased interest since the field of quantum information has come into existence. Heisenberg also held some anti-realist interpretation that would come under the participatory realist category today. The participatory realistic interpretations of the quantum theory include the Copenhagen interpretation, QBism, and other various informational interpretations. The participatory realist interpretations do not view the probabilities of measurement outcomes of quantum theory as determined by the intrinsic properties of the system. Quantum theory does not deal directly with the intrinsic properties of the system but with experiences, an observer has of the system. Similar to statistical interpretation the quantum state is viewed epistemically but there is no deeper underlying reality as in the case of statistical interpretation. Quantum states carry an observer's knowledge about the results of future experiments, or `about belief' if they view the quantum state as an observer's expectations about the results of future actions. The quantum states in QBism are viewed as non-contextual beliefs about the system. QBists call the interpretation $\psi$-doxastic. The properties are `doxxed' after measurement. Quantum mechanics provides the observer with conditions for assigning probabilities via Gleason's theorem. The QBists have so far managed to evade every problem thrown at them. Though we find QBism a bit too subjective for our liking it seems to be interesting nevertheless.

Objectivity has always been the necessary feature of physical theories. The recent works on epistemic interpretations of quantum states pose the threat of loss of objectivity in physics. QBists think this is a misconception and try to reintroduce the observer back. What we have to acknowledge is that while participatory realist interpretations might provide alternatives to ontic interpretation and statistical interpretation, the problems with quantum evolution persist. Some changes have to be made in quantum evolution to fix the problem.

\subsubsection{The Observer}\label{section:observer}
One common feature among all the interpretations belonging to the participatory realism category is their acceptance of objects of quantum theory along with the collapse rule. Many of the participatory realist interpretations have some important role for observers or agents. Physics has always been about describing the world without getting into discussions about observers. To describe observer or agent we have to rely on some pre-theory that's not the quantum theory itself and this can lead us into a never-ending, unproductive philosophical discussion. By using the instrumentalist approach Ludwig tried to remove this dependency of quantum theory on pre-theories. We will try to describe the notion of observer in the instrumentalist lingo. One of Ludwig's main motivations was to show that quantum mechanics, with all its subtleties, can be formulated and interpreted consistently without radical changes in our concept of physical reality and to reject the decisive role of human consciousness in the creation of observable events. Now we are in an awkward situation where we intend to describe the notion of observer using Ludwig's lingo. 

Observers have always been present in physical theories. We never had to describe the `observers' own state space and things of that sort in classical physics. Observers are extremely important in relativity as well. This situation doesn't change in the case of quantum mechanics. Observables make sense only in the presence of observers. Without observers, there is no meaning in saying that an observable has so and so value. It's the so and so value observed by some observer. The observations are described by the devices used for the measurements. These instruments and events are our links to `objective reality'. The description of effects in quantum theory already contains the notion of observer. The mathematical structure of effects already contains all the structures needed for the mathematical description of observer. The observer together with the measuring device can (should?) be identified with the effects of the system. 
$$\text{Observer}\hookrightarrow \text{Effects}$$
In the case of quantum theory, we identify the observer with the effects that the observer can measure. 
$$\text{Observer}\hookrightarrow \mathcal{P}(\mathcal{H})$$
where the lattice of projection operators, $\mathcal{P}(\mathcal{H})$ acts as the space of effects.

This allows us to introduce the notion of observer without having to introduce state space for observers and mutilating quantum theory. Observers are implicit in the formulation of quantum theory. The problem of pre-theories is evaded. The information the observer has about the system is contained in the states of the system. In this sense, there is nothing special about the role of an observer in quantum theories when compared to classical theories. Closely related are the works of Bartlett, Rudolph, and Spekkens, \cite{Bartlett} and the more recent work of Krumm, H\"ohn and M\"uller \cite{Krumm} and the references therein. The keyword is structuralist approach to reference frames. By observer, we don't mean a human or a conscious being. We would like to stay agnostic regarding these issues. We feel these issues are not of concern to physicists. The universe will exist even if there are no living beings. The good old Copenhagen interpretation of quantum states might be perfectly fine for our purposes. It's not important to discuss which of the many participatory realist interpretations is better, most of them will be compatible with what we wish to do in the next chapter.
\begin{center}
\end{center}


\chapter{\textsc{Guidance from Thermodynamics}}
This chapter aims to seek guidance from thermodynamics to guess a law of evolution for quantum theory. In particular, we are interested in the notion of time and the constraints on evolution. For that purpose, it's important to study the relation between quantum theory and thermodynamics. 

Physics does not consist of one theory. It's however a common belief that there should be one theory behind all these physical theories. We will say two theories are compatible if it's possible to go from the objects and laws of one theory to the other. We expect two theories with the same application domain to be compatible. If the application domain of one theory contains the other then it should be possible to approximate the other using the first theory. Suppose we have a class of physical theories $\{PT_\alpha\}_{\alpha\in I}$,
$$PT_\alpha\equiv A_\alpha \longleftrightarrow MT_\alpha,$$
where $A_\alpha$ is the application domain and $MT_\alpha$ is the corresponding mathematical theory. The correspondence rules associate with facts in $A_\alpha$ physical objects in $MT_\alpha$ and the laws of the theory provide morphisms between these objects. We are interested in comparing physical theories.  Suppose $PT_\alpha$ and $PT_\beta$ are physical theories with the same application domain $A_\alpha \equiv A_\beta$. Now each theory will have its own physical objects, idealization process, and correspondence rules. The two theories are the same if all these are the same. These will be different forms of the same theory. We are interested in a relation of the form, $PT_\alpha$  is better than $PT_\beta$ for $\alpha, \beta\in I$,
$$PT_\alpha \succ PT_\beta,$$
that is, $PT_\alpha$ says more about the structure of $A_\alpha\equiv A_\beta$ than $PT_\beta$. {All these can be made precise, see \cite{Ludwig}.} In such a case there must exist some kind of forgetful map,
$$PT_\alpha \xlongrightarrow{F} PT_\beta.$$ 
We expect these maps to map objects of $PT_\alpha$ to objects of $PT_\beta$ and laws of $PT_\alpha$ to laws of $PT_\beta$. The maps between physical theories need not necessarily be forgetful maps, they can be approximations like in the case of quantum to classical maps or some other general map. But they should take objects of the physical theory to objects and laws to laws. Physical theories will be categories and these maps will be functors. The theories in question are quantum mechanics $\mathcal{QM}$ and thermodynamics $\mathcal{T}$. We expect both thermodynamics and quantum mechanics to be universal and we expect quantum mechanics to be better than thermodynamics.

Before quantum physics, classical mechanics was considered universal. Statistical mechanics provides a connection between classical mechanics and thermodynamics. Denote classical mechanics by $\mathcal{CM}$. Statistical mechanics is the map,
$$\mathcal{CM}\xlongrightarrow{F_C}\mathcal{T}.$$
Since quantum mechanics should replace classical mechanics as an attempt at a universal theory we should expect a statistical mechanics type connection between quantum mechanics and thermodynamics.
$$\mathcal{QM}\xlongrightarrow{F_Q} \mathcal{T}.$$
What we want to do is guess the laws of $\mathcal{QM}$ intelligently. We know $F_Q$ should exist. We know what $\mathcal{T}$ is, at least vaguely. The aim is to find a law for the quantum theory that would satisfy the above relation. The thing we cannot do is statistical mechanics like procedure for the quantum case because probabilities in quantum mechanics cannot be viewed statistically. 

\section{Laws of Thermodynamics}
It's important to note what the laws of thermodynamics say and what they don't. Thermodynamics is the study of thermal equilibrium. Its state space is the space of equilibrium states of the system. These equilibrium states, by definition, do not evolve if left to themselves. The states of thermodynamical systems change because we do things to them. The transitions between the equilibrium states are called thermodynamic processes. These processes are a result of an outside intervention under a set of control variables. Thermodynamic laws are different from other physical theories. Thermodynamic laws don't say how systems evolve from one state to other. The laws of thermodynamics describe the properties these equilibrium states have, their relation with control variables, and thermodynamic processes. The zeroth law expresses the transitivity of thermal equilibrium. It also implies the existence of temperature as a parametrization of equilibrium states. The first law introduces the notion of energy. It relates mechanical work and the notion of heat using the notion of energy for systems at equilibrium. The existence of the special variable called energy and the fact that the amount of energy in an equilibrium state is independent of how the state was arrived at is the first law of thermodynamics. It provides a connection between mechanics and thermodynamics. It allows the notion of energy to be used as one of the parameters describing the equilibrium states. The different formulations of the second law tell us what kind of transitions of the system are not allowed to happen. The second law leads us to the existence of an entropy function which tells us which processes can occur and which cannot. The second law from purely classical thermodynamics sense does not talk about the trend of non-equilibrium states towards equilibrium states.

There are various axiomatizations of thermodynamics. Depending on what they choose as fundamental each author has their own way of axiomatization. Here we will give a brief sketch of Lieb \& Yngvason's approach to second law \cite{Lieb1}, and go into a discussion on certain topics that are of interest to us, especially those related to the notion to time \cite{Uffink1}.

\subsection{Is the Second Law Relevent? (Probably not)}
A system satisfies the second law of thermodynamics if there exists a function $J$ that tells us which processes are allowed. Lieb \& Yngvason highlight some basic intuitive assumptions classical systems satisfy which guarantee that the system will satisfy the second law. Starting from a few basic assumptions they explicitly construct an entropy function.

The equilibrium states correspond to points in the space of states $\Gamma$ of the thermodynamic system. In the next step, the space $\Gamma$ is given the structure such that one variable is the energy $U$ and the remaining variables come from some pre-theory. It's at this stage the notion of energy enters physics. In classical thermodynamics, the variables coming from pre-theory were volume, pressure, etc. The space of states $\Gamma$ is taken to have the structure of a convex set in $\mathbb{R}^{2n+1}$, where one variable is energy $U$ and the remaining are mechanical variables.
The scaling of the system $\Gamma$ by $\lambda\in\mathbb{R}^+$ will be denoted by $\Gamma^\lambda$.\footnote{This scaling will not be possible in the quantum case.} For $\lambda,\mu\in\mathbb{R}^+$, we should expect $(\Gamma^\lambda)^\mu=\Gamma^{\lambda\mu}$. Two systems $\Gamma_1$ and $\Gamma_2$ can be combined and the states are points from the cartesian product, $\Gamma_1\times\Gamma_2$.

A state $\sigma$ is called adiabatically accessible from $\rho$ if there exists a thermodynamic process that takes $\sigma$ to $\rho$. We denote it as,
$$\rho\prec \sigma.$$
This relation $\prec$ is intended to introduce some ordering on the space of states $\Gamma$. A process is irreversible adiabatic if $\rho\prec \sigma$ and not $\sigma\prec \rho$ we denote this by $\rho\prec\prec \sigma$. Two states $\rho$ and $\sigma$ are adiabatically equivalent if $\rho\prec \sigma$ and $\sigma\prec \rho$ and we denote it by $\rho\sim \sigma$. 

Lieb \& Yngvason list some basic requirements on this relation $\prec$ that guarantees the existence of entropy function, $J:\Gamma\to\mathbb{R}$. The trivial process that takes $\rho$ to $\rho$,
$$\rho\sim \rho.$$
If we divide the system into $\Gamma^\lambda$ and $\Gamma^{1-\lambda}$ and then combine the two, the resulting system should be equivalent to the original. 
$$\rho\sim (\lambda \rho,(1-\lambda )\rho),\:\:\:\:\:\:\lambda\in[0,1].$$
If $\rho\prec \sigma$ and we scale both the systems by $\lambda$ then there should be a process that takes $\lambda \rho$ to $\lambda \sigma$. If there is a process that takes $\rho$ to $\sigma$ and a process that takes $\sigma$ to $\gamma$ then we can go from $\rho$ to $\gamma$ by doing the first process and then doing the second process. The relation $\prec$ is transitive.
$$\rho\prec \sigma, \lambda > 0 \:\:\implies \lambda \rho\prec \lambda \sigma.$$
$$\rho\prec \sigma \:,\: \sigma\prec \gamma \:\implies \:\rho\prec \gamma.$$
If there exists a process such that $\rho\prec \rho'$ in $\Gamma_1$ and $\sigma\prec \sigma'$ in $\Gamma_2$ then we can go from $(\rho,\sigma)$ to $(\rho',\sigma')$ in $\Gamma_1\times \Gamma_2$. 
$$\rho\prec \rho', \sigma\prec\sigma' \implies (\rho,\sigma)\prec (\rho',\sigma').$$
We expect small perturbations to not change the ordering of two systems. 
$$(\rho,\epsilon \gamma)\prec (\sigma,\epsilon \gamma'),\epsilon \to 0  \implies \rho\prec \sigma.$$
For every state $\rho\in\Gamma$ there exists a state $\sigma$ such that $\rho\prec\prec \sigma$. This is a crucial assumption about $\Gamma$.
$$\rho\in\Gamma \implies \exists \:\:\sigma \text{ such that } \rho\prec\prec \sigma.$$
Two states are said to be comparable if either $\rho\prec \sigma$ or $\sigma\prec \rho$. A state-space $\Gamma$ is said to satisfy the comparison hypothesis if all pairs of states in $\Gamma$ are comparable. For a state-space $\Gamma$ satisfying the above order relations and the comparison hypothesis one can construct an entropy function as follows. The comparison hypothesis can be derived from a few more basic properties we expect from systems at equilibrium and thermal contact but we are not interested in that.

Let $\Gamma$ satisfy the comparison hypothesis, choose some base states $\rho_0$ and $\rho_1$. For any given state $\rho$, since every elements of $\Gamma$ are comparable, we have for all $\lambda>0$ we have either $\rho\prec((1-\lambda)\rho_0,\lambda \rho_1)$ or $ ((1-\lambda)\rho_0,\lambda \rho_1)\prec \rho$. This is possible because the state $((1-\lambda)\rho_0,\lambda \rho_1)$ is comparable to $((1-\lambda) \rho,\lambda \rho)\in \Gamma^{1-\lambda}\times \Gamma^\lambda$ which is equivalent to $\rho$. Then define $J(\rho)$ as,
$$J(\rho):=\sup\{\lambda\:\:|\:\:((1-\lambda)\rho_0,\lambda \rho_1)\prec \rho\}.$$
It's easier to think about this by taking some example. Take water for example, we can take $\rho_0$ and $\rho_1$ to be ice and boiling water, and given $1$ amount of water at $\rho$, $\lambda$ is the amount of boiling water needed to convert $(1-\lambda)$ amount of ice into $1$ amount of water in $\rho$. It can be checked that the function $J:\Gamma\to \mathbb{R}$ has the following properties, 
$$\rho\prec \sigma \implies J(\rho)\leq J(\sigma).$$
\vspace{-1.7em}
$$J((\rho,\sigma))=J(\rho)+J(\sigma).$$
$$J(\lambda \rho)=\lambda J(\rho).$$
This function serves as our entropy function. If we change the base points the change will be of the form,
$$J(\rho)\to aJ(\rho)+B.$$
Assume that the comparison hypothesis holds for all compound systems. For each system $\Gamma$ let $J_\Gamma$ be some definite entropy function on $\Gamma$. It's possible to choose constants $a_\Gamma$ and $B(\Gamma)$ such that the function $J$ is defined for all states of all systems by,
$$J(\rho) = a_\Gamma J_\Gamma (\rho) + B(\Gamma).$$
What entropy allows us to do is tell which transitions are possible. If $\rho\prec \rho'$ and $\sigma\prec \sigma'$ then we already know $(\rho,\sigma)\prec (\rho',\sigma')$, entropy tells us nothing new about such a system. $J(\rho)+J(\sigma)<J(\rho')+J(\sigma')$ doesn't imply $J(\rho)<J(\rho')$. The compound systems can have a process $(\rho,\sigma)\prec (\rho',\sigma')$ but $\rho\nprec \rho'$. The inequality $J(\rho)+J(\sigma)<J(\rho')+J(\sigma')$ tells us exactly which adiabatic processes are allowed independent of any data about the manner in which the systems interact. A system with the $\prec$ satisfying the above conditions will have the second law. The existence of such a function $J$ is called the entropy principle. 

The second law, in this view, refers to processes of a system that begin and end in equilibrium states and says that the entropy of the final state is never less than that of the initial state. The second law does not say anything about systems reaching an equilibrium state. The laws of thermodynamics don't say anything about the evolution of isolated systems. 

\subsection{The Equilibrium Principle (Relevent)}
It's a common belief among physicists that thermodynamics says something about the flow of time. These discussions often assume that the second law is where time asymmetry enters thermodynamics. Is it really the second law where time asymmetry enters? The starting point of most formulations of thermodynamics is equilibrium states which by definition don't change in time. The second law, after all, provides a relation between variables at equilibrium. The question we should be asking about the second law is regarding its scope. What's the scope of the second law of thermodynamics? If we take the definition of equilibrium seriously, and if the second law is concerned with transitions from one equilibrium state to another, is it applicable to the universe as a whole? so that we can say the universe's entropy is increasing or does it only apply to sub-systems of the universe? If it does apply to the whole universe, how does the universe go from one equilibrium state to another and why? Equilibrium states are by definition supposed to stay unchanged and the universe has nothing influencing it from the `outside'. Isn't it supposed to remain in the same state once attained? It is usually assumed that the approach to equilibrium is accompanied by an increase in entropy and that this is a consequence of the second law. Perhaps it could be extended to include non-equilibrium systems, but thermodynamics in its traditional sense doesn't say anything about systems trending towards equilibrium. The idea that the second law says something about the flow of time lacks a good theoretical foundation. The origins of the flow of time might be something deeper as pointed out in \cite{Uffink}.

There is however no question that thermodynamics, if not its second law, makes time-asymmetric claims. The spontaneous movement from non-equilibrium to equilibrium happens and is assumed throughout the field. This most certainly is a time asymmetric notion. This fact comes logically before the laws of thermodynamics. The existence of this tendency should not be confused with the zeroth law of thermodynamics which says about the transitivity of interbody thermal equilibrium. The tendency towards equilibrium is a more basic principle than that of transitivity. Brown, Uffink \cite{Uffink} call this the Equilibrium principle,

\vspace{1em}
\begin{postulate}
	{\bfseries\textsc{(Equilibrium Principle)}} An isolated system in any initial state will spontaneously attain a unique state of equilibrium.
\end{postulate}
\vspace{1em}

The equilibrium principle guarantees the existence of equilibrium states for isolated systems. These states by definition remain the same once they are attained. The equilibrium principle tells us about the uniqueness of the equilibrium state; i.e. for any initial state of an isolated system, there is exactly one state of equilibrium to which the system will reach. This is where the time asymmetry arises at the most basic level. The existence of such states itself is a time-asymmetric notion. The time asymmetry is built into thermodynamics by way of the notion of equilibrium. Time asymmetry may not be needed for a statement of the second law but is definitely required in the interpretation of the notion of equilibrium. Instead of getting into a discussion about the existence or the non-existence of a connection between the second law of thermodynamics and the notion of time (which is a common naive opinion held by most physicists), we will focus on the equilibrium principle which we know for sure has a connection with the notion of time. Equilibrium principle is more relevent to our discussion than the second law of thermodynamics.

\section{Evolution in Quantum Theories, Revisited}
In this section, we try to guess the time evolution in the quantum case. The equilibrium principle is the only law of thermodynamics that's similar to the laws of other physical theories. It tells us about states going to other states like in other physical theories. It doesn't however say anything about how states are going to evolve. Our aim here is limited. It's to construct a quantum counterpart that is compatible with the equilibrium principle of thermodynamics. That is to say, this new construction would allow us to arrive at the notion of equilibrium from the quantum side. 

This task can be further divided into the following questions; how does a system evolve given the structure of observables and states? why that might be the case? what mathematical structure of the physical theory plays a crucial role in the said behavior of the system? We will start with the second and the third question first.

\subsection{Emergence of Time}\label{section:Time}
Time is a very basic physical quantity. Prerelativity physics had a very clear notion of time, the absolute time. It was believed that time flows without relations to anything intrinsic to the system. Initiated by Lorentz, Einstein, and others, the twentieth century questioned this notion of time for the first time. Quantum physics which was under development at the time however continued with the old notion of time. Time is absolute in quantum physics and not so in relativity. This conceptual difference is referred to as the problem of time. In classical physics, the notion of time is associated with the time translation symmetry of the system. The symmetries of the system are assumed apriori in quantum theory. The notion of time and other important physical objects in quantum theory comes from these pre-theories. This places quantum theory in an awkward situation of being a more fundamental theory that relies on an outdated theory. The distinction is between the variable `time' that appears in physical theories for instance in Newtonian physics and the `time' that flows in our experience. The first is well defined and is unproblematic by itself and is related to the symmetries of the system (as suggested by Rovelli, we can just forget about this notion of time \cite{Rovelli3}). This notion of time is reversible, in the sense that the dynamics of the physical processes remain well defined when the direction of time is reversed. This is in strong contradiction with our experience of time as we perceive it. It does not justify the sense of passage and flowing that we associate with time. So, it is the second that needs explanation. It should hence be stressed that we cannot be identifying physical time with the one-parameter group generated by the Hamiltonian. Any notion of `physical time' should come equipped with a few peculiar features. These features should include, the distinction between various instants of time, the existence of a notion of time instances close to each other, the possibility of arranging the instants of time on a one-dimensional manifold, a metric structure, an order, and a notion of `now'. What we are interested in is the emergence of such a notion of time in quantum theory.

Consider two observables $A$ and $B$ that are being measured consecutively. If the system is in the state $\rho$ and the outcome of the measurement for the observable $B$ is $B_j$ then the state after the measurement will be given by, $$\rho_{B_j}=\frac{E_{B_j}\rho E_{B_j}}{Tr[E_{B_j}\rho E_{B_j}]}.$$ Suppose the observable $A$ is measured after the above measurement then the probability that the observable $A$ takes the value $A_i$ will be given by $p(A_i|B_j)=Tr[\rho_{B_j}E_{A_i}]$. If the measurement gives the value $A_i$ then the state of the system will be, $$\rho_{A_iB_j}=\frac{E_{A_i}E_{B_j}\rho E_{A_i}E_{B_j}}{Tr[E_{A_i}E_{B_j}\rho E_{A_i}E_{B_j}]}.$$ If the two observables $A$ and $B$ are compatible then $\rho_{A_iB_j}=\rho_{B_jA_i}$. Suppose $A$ and $B$ are compatible then the measurement sequence $A\to B\to A$ is equivalent to the measurement sequence $A\to A\to B$. Hence in the first sequence of measurements, the value of the observable $A$ obtained in the last measurement will be the same as the value in the first measurement. Commuting observables can be measured so that the values of all the observables can be known at once, in the sense that if you conduct a measurement of either observable $A$ or $B$, then it will be possible to say with certainty what the values are going to be. Since the algebra of observables is non-commutative, having complete knowledge about the observables is not possible. The system can always be measured to get new information. Non-commutativity forces the system to always have some information about it and the observer\footnote{By observer, we mean in the sense described in the previous chapter, \S\ref{section:observer}. We don't mean a human or any conscious being. } can keep obtaining new information from it. This is the reason for irreversibility in quantum theory. We will call a process by which an observer updates their knowledge about the system an observation. An observation need not necessarily change the state, for example, measuring the same observable again won't give any new information. 

Perception of time is an elementary experience. Heuristically, time changes for an observer if the observer notices some change in the system. If the observer notices no change at all then it's equivalent to no change in time. So perhaps it's the other way round? When a system is measured for an observable, before the measurement the value of the observable is unknown, and after the measurement, we know the value of the observable. Every quantum process brings with it a notion of before and after. So it might not be that if the observer notices no changes in the system is equivalent to time not changing, but that it's \textbf{because} of the observations time changes for the system. Every quantum process changing the state of the system should correspond to some time change and conversely, to each change in time, there is some change in the state. 

\vspace{1em}
\begin{postulate}
	{\bfseries\textsc{(Time)}} Time changes for a system \textbf{because} of quantum processes.
\end{postulate}
\vspace{1em}

This can be stated in a more clickbaity way as, Observation of the system changes time for the observer. The purpose of using the term observation instead of terms like process or event is that it might provide some intuition to the reader. This tells us why the notion of time even exists. The mathematical structure responsible for this is the non-commutativity of observables in quantum theory. It's the non-commutativity of the observables that allows us to arrive at this notion of time. Any generalized probabilistic theory that has some similar mathematical structure will possess this notion of time. This notion of time cannot exist in the commutative case, for example, say classical physics. This is closely related to the notion of `events' in Haag's work, \cite{Haag2}. Jürg Frohlich and collaborators at ETH have an interpretation which they call Events-Trees-Histories interpretation or cleverly abbreviated as the `ETH interpretation' \cite{BlanchardFrohlichSchubnel}, which seem to be closely related to Haag's ideas and hence the ideas presented here. However in their approach time is an irreducible concept and events monitor time. In our case time is an emergent concept and events are the reason for the existence of time. The notion of time exists just because of the structure of observables and no additional structure is required. 

The notion of time as emerging out of random observations/events doesn't give any structure to it. We don't have any additional structure that gives us the `flow' of time yet. This is contradictory to our experience. We expect time to have some additional structure. Our next step is to understand how this structure could come into quantum theory when our trees of events have some additional constraints. We will then relate it with the already existing notions of time in mechanics, thermodynamics, etc.

\subsection{Approach to Equilibrium}
Everything we have done so far on the quantum side is subjective to the observer. This is however in contradiction with common sense that there must be certain things that are objective in the physical world. `{How come one world out of many observer participants?}' was one of Wheeler's three questions in \cite{Wheeler}. 
\subsubsection*{Wigner's Friend Type Thought Experiment}
One way to bring objectivity into quantum theory is through intersubjectivity. To describe how objectivity might come into the quantum picture consider the following Wigner's friend type situation. Before discussing this however we will again visit the notion of observer and what constraints we have on modelling observers. In \S\ref{section:observer}, we gave reasons as to why we should not treat observers as having their own state space. If observers are themselves taken to be quantum systems, with their own state space and all them we get into some inconsistencies like in the Frauchinger-Renner's paper \cite{FrauchingerRenner}. Though this inconsistency might be coming from neglecting the collapse rule, we feel assigning to observers their own state space should also be problematic. Interested reader should check out \cite{Bub} \& \cite{Bubsingle} for a discussion on this. Identifying observers with the algebra of observables is a very minimalist approach and doesn't mutate quantum theory.

Suppose Wigner's friend performs a measurement in her laboratory and obtains an outcome that Wigner who is not in the laboratory doesn't know. The quantum state for Wigner gets updated once his friend tells him the result of the experiment. Here Wigner's measuring apparatus includes his friend. If the two performed the experiments together, they have to agree with the result of the experiment. We would like to extend this thought experiment further. Wigner performs an experiment in one room and his friend performs some other uncorrelated experiment in a different room, they meet and tell each other the results of their experiment. Now for each of them, their state updating is different. Though they will eventually end up agreeing about the results of the experiments, the way the state updating happens, i.e, the way time changes, for each of them is different. However, the information they have about the experiments will be the same in the end. If multiple agents are observing the system, they should eventually come to terms with their knowledge of the observables.

Objectivity only makes sense if the observers are observing the same observables. Let the first observer be represented by $(\mathcal{A},\rho)$ and the second observer by $(\mathcal{B},\sigma)$, where the first term is the algebra of observables and the second term is the state which corresponds to the observer's knowledge about the observables. To make sense of objectivity the algebra of observables for both observers should have some intersection and the observable being measured should be in this intersection. After measuring an observable $A\in\mathcal{A}\cap \mathcal{B}$, the result is the same for both the observers if the restriction of their respective states to the subalgebra generated by the operator $A$ coincide. If we denote the algebra generated by $A$ as $\mathcal{A}_A$ we expect,
$$\rho|_{\mathcal{A}_A}=\sigma|_{\mathcal{A}_A}.$$
If this happens we can say that the information the two observers have about the observable described by the operator $A$ is the same. We are interested in observers who are observing the same observables $\mathcal{A}=\mathcal{B}$. Heuristically, the intersubjectivity should force the state evolution to be such that they will eventually reach some common state. There is no way to know this class of objective states of the system if only the algebra of observables is known. The existence of such states is an additional physical hypothesis.

\vspace{1em}
\begin{postulate}
	{\bfseries\textsc{(Objectivity)}} For every system there exists a class of objective states $\mathfrak{O}$.
\end{postulate}
\vspace{1em}

This, in some way, answers the question, why does a system evolve? The vague answer would be, because it has to reach one of the objective states. The class of objective states should form the building blocks for everything objective we see in quantum theory including the laws of evolution. The purpose of the laws of physics is to recognize some regularity in naturally occurring processes that are in general complicated and unpredictable. These regularities are then formulated as laws of physics. What we have now is a collection of events $\mathfrak{T}$ which we will call a tree (to be close to other works in foundations of quantum physics), that take the system towards some objective state in $\mathfrak{O}$. Which of these trees is realised can only be determined through Bayesian inference (updating the state based on the information available to us). What we want to be able to do is assign to each event in a tree a `time'. In order to do that we must first define some properties the class of objective states should have, then use these properties to define to each event a `time' that coincides with our intuition regarding the notion of time. 

\subsubsection{Relation to Thermal Time Hypothesis}\label{section:thermal}
Heuristically, the class of objective states should be closely related to the equilibrium states. 
Equilibrium states are closely related to Hamiltonian of the system. The first law, in the sense of \cite{Lieb1}, says that energy is a coordinate in the description of equilibrium states.

\begin{postulate}
		{\bfseries\textsc{(First law of Thermdynamics)}} Energy is a coordinate for the description of equilibrium states.
\end{postulate}

Let the operator corresponding to energy is $H$, Jaynes and Bayer \& Ochs' work \cite{Jaynes1}, \cite{BayerOchs} show that equilibrium states maximize entropy under the constraint of expectation value of energy.
This postulate relates the class of objective states and the Hamiltonian of the system. Given the objective states, the thermal time hypothesis allows us to recover the Hamiltonian. The notion of thermal time was put forward by Connes and Rovelli \cite{carlotime}. The thermal time of a system is the natural flow induced by its state on its algebra of observables. So far, however, thermal time has remained a rather abstract notion with a few concrete applications \cite{Rovelli4}. 

For a finite system $\mathcal{A}$ with the Hamiltonian $H$, the time translation symmetry is given by, $\alpha_t=e^{-iHt}(\cdot)e^{iHt}$. The equilibrium state at temperature $1/\beta$ is characterized by the Gibbs' condition given by,
$$\rho^\beta(A)= \frac{Tr(e^{-\beta H}A)}{Tr(e^{-\beta H})}.$$ 
For the case of any general system, the KMS condition given by,
$$\rho^\beta(A\alpha_{i\beta}(B))=\rho^\beta(BA).$$
characterizes the equilibrium states. 

For a von Neumann algebra $\mathcal{A}$ acting on a Hilbert space $\mathcal{H}$, state $\rho=\langle \Omega_\rho |\cdot \Omega_\rho \rangle$, we have the embedding of $\mathcal{A}$ in $\mathcal{H}$ via the map, $A \mapsto A\Omega_\rho$, the $^*$ operation on $\mathcal{A}$ gives an anti-linear operator $S$ on $\mathcal{H}$,

\[\begin{tikzcd}
	\mathcal{A} \arrow{rr}{A\mapsto A^*} \arrow[swap]{d}{A\mapsto A\Omega_\rho} && \mathcal{\mathcal{A}} \arrow{d}{A\mapsto A\Omega_\rho} \\
	\mathcal{H}\arrow{rr}{S:\:A\Omega_\rho\mapsto A^*\Omega_\rho}	&&\mathcal{H}
\end{tikzcd}\]

\vspace{1em}
\noindent Then, $\Delta=S^* S$ is a positive self-adjoint operator. The Tomita modular operator $\log \Delta$ implements a one-parameter automorphism group $t\mapsto\alpha^\rho_t$ of $\mathcal{A}$ where,
$$\alpha^\rho_t(A)=e^{it\log \Delta}Ae^{-it\log \Delta}.$$
called modular automorphism. The parameter $t$ is called modular time. Tomita-Takesaki theorem says that the state $\rho$ satisfies with KMS condition with respect to the flow $\alpha^\rho_t$ with inverse temperature $\beta=-1$. Given the time translation, one can determine the equilibrium states by the KMS condition, and assuming a state is in equilibrium one can construct the one-parameter group of automorphisms with respect to which the state satisfies the KMS condition. 
\subsubsection{Why not standard Unitary evolution?}
Given a state $\rho$ of the system $\mathcal{A}$, the thermal time flow $\alpha^\rho_{-t}$ is the modular flow associated with the system $\mathcal{A}$. For more on modular theory, see \cite{Longotime},\cite{Summers}. The problem however is that the thermal time hypothesis associates to every state a flow. Not all these flows have physical meaning. Random states don't correspond to physically meaningful flows. The modular group cannot be directly related to the flow of time. The thermal time hypothesis does however tell us something about the notion of time and its relation to time. There is a need to explain why certain states are special compared to others. So the question then is what is the connection between the modular theory of operator algebras and thermodynamics? What it allows us to do is construct the Hamiltonian. This is similar to what the first law says, that energy is a variable for the description of equilibrium states. What modular theory and the thermal time hypothesis allow us to do is construct this variable from the equilibrium states. The equilibrium states contain data regarding the Hamiltonian of the system and by Tomita-Takesaki theory this data can be recovered. Starting from the Hamiltonian of the system one obtains the equilibrium states by the KMS condition and conversely assuming a state is an equilibrium state one can construct the Hamiltonian. One can then ask what is the difference between these two? Does using equilibrium states as starting point add any new methods to physics? If the two were the same then this approach wouldn't add anything new to physics. It's important to clarify the difference between the approach where Schr\"odinger equation is the starting point and the approach where the objective states corresponding to equilibrium states are the starting point. Schr\"odinger equation is a constraint on the evolution of the system. It tells us that systems must follow the unitary evolution, coming from the symmetries of the system. If our starting point is the objective states, the constraint is that the system trends to the objective states. It is not as strict a constraint as Schr\"odinger equation. The disadvantage of the Schr\"odinger equation is the following, if the constraints coming from Schr\"odinger equation were true, it would mean an isolated system would never approach equilibrium. We expect this to be not the case. However, if the constraints are relaxed a bit and we postulate only that the systems trend to equilibrium then we have a lot more flexibility and many more physically observed phenomenon can be explained while also recovering the information regarding symmetries through modular theory. The only difference between the two is the starting point, in the standard case, the symmetries of the system are the starting point, in our case the equilibrium states are the starting point. Both the approaches carry the same information about the system but evolution in the standard approach is more constrained than our approach.

In-order to explain why equilibrium states are special out of all the states, we unfortunately (fortunately?) have to invoke the maximum entropy-minimum relative entropy methods for now and hope some better explanation appears in the future.

\subsection{Arrow of Time}
The processes of interest to us are those that take the state closer to an objective state. Our next step should be to define a notion of `time' for systems tending to equilibrium. From a pragmatic point of view, the dynamics of the system should be about studying how the system tends to an objective state. Since the objective states are invariant under time-translation symmetry of the system, this boils down to studying the loss of asymmetry. To make quantitative statements regarding this, we rely on the resource theory of asymmetry. 
\subsubsection{Resource Theory of Asymmetry}
A resource theory consists of a subset $\mathcal{F}$ of all quantum channels called free operations, closed under composition, contains the identity and a set of free states $\mathcal{S}$. Any states that cannot be created from free states by performing the free operations is called a resource state. If $\rho$ is a resource state and some other state $\sigma$ can be reached by performing a free operation on $\rho$ then the state $\rho$ is a better resource than $\sigma$.
$$ \alpha(\rho)=\sigma \iff\rho \succ\sigma,$$
where $\alpha\in\mathcal{F}$. If the two states can be interconverted i.e, $\rho\succ\sigma$ and $\sigma\succ\rho$ then they lie in the same equivalence class $[\:\rho\:]$. The above-defined relation gives a partial order on the equivalence classes. Since it's not possible to convert a resource state into a free state using free operations alone we are interested in is what kind of quantum operations are needed to convert a resource state to a free state. The constraints on such operations will come from physical requirements.

Resource theory of asymmetry is a framework for quantifying the asymmetry of states and operations. Suppose we have a unitary representation of a group $\mathcal{G}$, $g\mapsto U(g)$ on the Hilbert space $\mathcal{H}$, free states of the resource theory of asymmetry of the group $\mathcal{G}$ are states with no asymmetry,
$$U(g)\:\rho \:U^{-1}(g)=\rho,\:\:\:\forall \:g\in \mathcal{G}.$$
We denote the totality of such states by $\mathcal{S}_{\mathcal{G}}$. A quantum channel $\alpha$ is a free operation if,
$$U(g)\:\alpha(\rho)\:U^{-1}(g)=\alpha(U(g)\:\rho \:U^{-1}(g)),\:\:\:\forall \:g\in \mathcal{G}.$$
We denote the set of all free operations by $\mathcal{F}_{\mathcal{G}}$. In order to study asymmetry of states, we first need a measure of asymmetry. A function $f:\mathcal{S}(\mathcal{H})\to \mathbb{R}$ is a measure of asymmetry with respect to the symmetry group $\mathcal{G}$ if it satisfies the following conditions,
$$f(\alpha(\rho))\leq f(\rho),\:\:\:\forall\: \alpha\in\mathcal{F}_{\mathcal{G}}.$$
$$f(\rho)=0,\:\:\:\forall \:\rho\in \mathcal{S}_{\mathcal{G}}.$$
The first condition says that allowed operations can only make the system more symmetric. The second convention fixes the value for symmetric states and makes the function non-negative. If the system is acted upon by an operation that takes it closer to an objective state, then the state must lose asymmetry in the process. The measures of asymmetry can be used to measure this change in asymmetry. Any changes in the asymmetry measure would indicate some change in the system. The measures of asymmetry can be used to parametrize this notion of time.
\subsubsection{A Description of Time}
Every event changes the state through some allowed operation and hence the state before the event and the state after the event will have different associated asymmetry. We expect the notion of time to have a few properties. For a system tending to an objective state or an equilibrium state, asymmetry measures can be used to distinguish different instants of time. Asymmetry measures map states onto real numbers, so to each state, one can associate a real number. The notion of time inherits an order and hence topological structure from these maps. As discussed before every observation brings with it a notion of before and after. This is closely related to a preferred instance of time representing the `now'. Suppose the system has followed a tree of events, $\mathcal{T}$, we can associate to each state $\rho\in\mathcal{T}$, its associated asymmetry $f(\rho)$. We can define the associated `time' for the state $\rho$ to be, $$t(\rho)=1/f(\rho).$$
This number captures most of our intuitive idea of the notion of `time'. We cannot however recover any metric structure for this notion of time. It's an a priori assumption that time is comparable from start of the universe to its end. The question one could ask then is the following, Is the flow of time in the early universe really comparable to the flow of time now? Is such an assumption necessary? 

\subsubsection{Relation to Mechanical Time}
The unitary evolution is not going to survive in a quantum theory of gravity. The simplest reason is that energy is not conserved in general relativity, and the standard quantum mechanics version of unitary evolution would keep the energy of the system constant. This is one of the first incompatibility between general relativity and quantum mechanics. One could find some workarounds like varying Hamiltonian but that would defeat the purpose of introduction of unitary evolution which was to implement the symmetries of the system. A replacement for unitary evolution is inevitable. The unitary evolution however, is a very good approximation for most purposes. So, it is important to relate the unitary evolution to the notion of time we just described. 

What we are interested in is the relation between dynamics and symmetries. That is, a relation between the notion of time we have here and the standard textbook version of evolution. For $\epsilon>0$ define,
$$\tau_\epsilon(\rho)=\inf \{s\:| \:J(\rho,\:e^{-isH}\rho \:e^{isH})\geq \epsilon\},$$ 
and is infinity if for all $s$, $J(\rho,\:e^{-isH}\rho \:e^{isH})$ is less than $\epsilon$. Here $J(\cdot,\cdot)$ is the relative entropy. As shown in \cite{MarvianSpekkens1}, for a `time translation'\footnote{Though we are using the traditional language we don't really identify it as such} invariant quantum channel $\alpha$, using the monotonicity of the measure of distinguishability and the commutativity of $\alpha$ with `time translation' we get,
\begin{align*}
	\tau_\epsilon(\alpha(\rho))&=\inf \{s\:| \:J(\alpha(\rho),(e^{-isH}\alpha(\rho)e^{isH}))\geq\epsilon\}\\&=\inf \{s\:| \:J(\alpha(\rho),\alpha(e^{-isH}\rho \:e^{isH}))\geq\epsilon\}\\
	&\geq \inf\{s\:|\:J(\rho,\rho(s))\geq\epsilon\}\\&=\tau_\epsilon(\rho).
\end{align*}
So we have,
$$1/\tau_\epsilon(\alpha(\rho))\leq 1/\tau_\epsilon(\rho).$$
Since symmetric states are invariant under the transformation $\rho \mapsto\:e^{-isH}\rho e^{isH}$, $1/\tau_\epsilon(\rho)$ will be zero in that case for all $\epsilon>0$. The function $f=1/\tau_\epsilon$ is a measure of asymmetry. So the standard unitary evolution is indeed related to our definition of time. This is applicable to any measure of distinguishability, $D:\mathcal{S}(\mathcal{H})\times \mathcal{S}(\mathcal{H})\to \mathbb{R}$ that only decreases the distinguishability between pairs of states for any quantum operation $\alpha$,
$$D(\alpha(\rho),\alpha(\sigma))\leq D(\rho,\sigma).$$
If the two states are identical then the measure of distinguishability should vanish.
$$D(\rho,\rho)=0.$$
Define $\tau_\epsilon^D(\rho)=\inf \{s\:| \: D(\rho,\:e^{-isH}\rho \:e^{isH})\geq \epsilon\}$, the function $1/\tau_\epsilon^D$ will be a measure of asymmetry. The one-parameter group generated by the Hamiltonian of the system though not the same as the flow of time is closely related to it. Depending on how the asymmetry measure changes, we can make claims about discreteness or continuity of the flow of time. We would like to stay agnostic about this.

It's important to note that the above quantification of the approach to equilibrium is unlikely the only way to study it. This approach reveals not much about the approach to equilibrium. The important point of the discussion is that we can define the notion of time based on the system's approach to equilibrium. What methods we use to quantify this approach to equilibrium is not important to us. Here we have used the invariance of the equilibrium states under the action of symmetry group. 

\subsubsection*{From Synthetic to Analytic?}
The description of time we have given so far is a synthetic one. It's described in terms of some inequalities and such. The physics community is usually adjusted to more analytic stuff, we usually work with objects that have nicer mathematical structures that allow us to do calculus for example. Reaching this stage however is complicated, one has to understand the properties of the object of interest well enough and those additional properties give us the necessary structure to develop analytic methods. We don't have anything to say as to how to reach this stage. While it's tempting to use Dilation theorems, we feel it's a faulty path to take (we expect quantum mechanics to have the capacity to explain closed systems as well). 

\clearpage
{\phantomsection\addcontentsline{toc}{chapter}{{Bibliography}}}

\end{document}